%% file: trans.tex
\documentclass[english]{article}

\input{defs.tex}
\input{notation.tex}
\begin{document}

\newcommand\authorstring{
Ian Holmes$^{1,\ast}$ \\
\textbf{1} Department of Bioengineering, University of California, Berkeley, CA, USA \\
$\ast$ E-mail: ihh@berkeley.edu
}

\newcommand\titlestring{Modular non-repeating codes for DNA storage}
\newcommand\shorttitlestring{Modular codes for DNA}
\markboth{\shorttitlestring}{\shorttitlestring}
\begin{flushleft}
{\Large \textbf{\titlestring} } \\
\authorstring
\end{flushleft}

\section{Abstract}
We describe a strategy for constructing codes for DNA-based information storage
by serial composition of weighted finite-state transducers.
The resulting state machines can integrate correction of substitution errors;
synchronization by interleaving watermark and periodic marker signals;
conversion from binary to ternary, quaternary or mixed-radix sequences via an efficient block code;
encoding into a DNA sequence that avoids homopolymer, dinucleotide, or trinucleotide runs and other short local repeats;
and detection/correction of errors (including local duplications, burst deletions, and substitutions)
that are characteristic of DNA sequencing technologies.
We present software implementing these codes, available at {\tt github.com/ihh/dnastore},
with simulation results demonstrating that the generated DNA is free of short repeats
and can be accurately decoded even in the presence of substitutions, short duplications and deletions.

\paragraph{Keywords:}
DNA storage, finite-state transducers, mixed-radix trees, deletion-insertion correcting codes

\tableofcontents

\section{Introduction}

DNA can store petabytes of information per gram \cite{GoldmanEtAl2013} and can last intact for tens of thousands of years \cite{GreenEtAl2010}.
This makes it an appealing prospect for long-term archival storage.
However, DNA synthesis, sequencing, and replication are prone to errors, which may limit its potential as a storage medium \cite{ChurchEtAl2012}.
These errors can be controlled by applying the tools of information theory,
treating DNA storage as a noisy channel coding problem.

Recently, several coding schemes for DNA storage have been proposed
that address the interrelated issues of error avoidance, error correction and redundancy
\cite{GoldmanEtAl2013,MilenkovicEtAl2005,YachieEtAl2008,YazdiEtAl2015,GuptaEtAl2015,MilenkovicEtAl2014,MilenkovicEtAl2015,GabrysEtAl2015,HuntEtAl2015,JainEtAl2016,MilenkovicEtAl2016,BornholtEtAl2016}.
Goldman {\em et al} \cite{GoldmanEtAl2013} made DNA using a ternary (radix-3) code that avoids repeated nucleotides,
synthesizing to fourfold redundancy for additional error correction.
Gupta {\em et al} introduced a ternary Golay code for error correction \cite{GuptaEtAl2015}.
Strauss {\em et al} used a Huffman code to map binary to ternary,
implemented a filesystem based on a random-access key-value store,
and improved on Goldman {\em et al}'s redundancy coding using an exclusive-or method \cite{BornholtEtAl2016}.
Milenkovic {\em et al} have considered
the avoidance of DNA structure in codes \cite{MilenkovicEtAl2005},
proposed several codes with analysis of their combinatoric properties \cite{MilenkovicEtAl2014,MilenkovicEtAl2015,GabrysEtAl2015,MilenkovicEtAl2016}
and presented an earlier design for a random-access rewritable filesystem \cite{YazdiEtAl2015}.
Avoidance of secondary structure was also discussed by Hunt {\em et al} \cite{HuntEtAl2015}.
Studies by Yachie {\em et al} \cite{YachieEtAl2008} and Jain {\em et al} \cite{JainEtAl2016}
have specifically addressed the storage of information in the DNA of living organisms,
the latter work (which appeared while this manuscript was in preparation) describing a duplication-correcting code.
Grass {\em et al} used a Reed-Solomon code for error correction \cite{GrassEtAl2015}.
Most such error-correcting DNA codes have focused on substitution errors, but there is recent
progress in the information theory field developing codes that correct insertions and deletions (``indels'').
For example, in prior work not specifically oriented to DNA storage,
MacKay {\em et al} developed synchronizing error-correcting codes that are resistant to
both substitutions and indels,
including watermark codes \cite{DaveyMackay2000,DaveyMackay2001}
and marker codes \cite{RatzerMackay2000}.
The Davey-MacKay watermark construction, in particular,
has been improved upon by other researchers (e.g. \cite{BriffaEtAl2010,JiaoEtAl2012,NguyenEtAl2013})
and even applied to DNA in the context of barcodes \cite{KrachtSchober2015},
though not for general DNA information storage.

Here we describe a modular strategy for constructing error-tolerant DNA codes.
As noted above, much recent work has used Goldman {\em et al}'s ternary code as a starting point.
The ternary code arises from avoidance of dinucleotide repeats, which leaves 3 available possibilities at each position; hence the radix of 3.
The rationale advanced by Goldman {\em et al} for this system is that it guards against the most common class of errors in DNA resequencing, which occur at homopolymer runs.
However, there are hidden assumptions in this reasoning; specifically, that it is worth spending a particular amount on synthesis
in order to mitigate a certain class of technology-dependent sequencing error.
Assuming that this is the case, it may also be desirable to avoid other error-prone motifs, such as repeated dinucleotides, trinucleotides or longer repeats.
Avoiding these sequences does not reduce the information content of DNA much, but they are not too far off homopolymer runs in their error rate \cite{LaehnemannEtAl2016}.
We may also wish to introduce error-correcting codes such as Hamming codes \cite{Mackay2003}, Golay codes \cite{GuptaEtAl2015},
turbo codes \cite{FreyMackay98,MurphyEtAl1999}, or Gallager codes \cite{Mackay1997}.
For entirely independent reasons, we might also want to constrain the coding scheme to avoid certain reserved codewords,
whether for biological function (e.g. binding sites for restriction enzymes or other factors)
or to organize our nucleic acid filesystem (e.g. barcodes, synchronization signals, or metadata block boundaries).
And yet none of this strategy can be taken for granted:
much of the above reasoning is predicated on the idea that it's worth mitigating error by synthesizing more DNA,
but (for the moment at least) synthesis is vastly more expensive than sequencing.
An alternate strategy would be to pack in as much information as possible and assume that sequencing will be cheap enough to correct errors.

In short, different application requirements and changing economics may require different codes.
As noted above, several researchers have developed solutions to some of these problems.
Here, we combine some of these ideas, and introduce some new ones, using a modular strategy for code design.
With this method, codes can be assembled to meet requirements
including error-avoidance, error-correction, and demarcation of metadata.

The core idea of our approach is to convert raw binary data into a mixed-radix sequence
wherein successive digits may be binary (radix-2), ternary (radix-3) or quaternary (radix-4).
The radix at any given position is effectively specified by the number of nucleotides available for coding at that position
(which may vary due to avoidance of repeats or reserved motifs).
The mixed-radix sequence can then be efficiently converted to a DNA sequence.
Error-correcting codes (for example, Hamming codes that introduce parity bits, or codes that maintain synchronization in the presence of indels)
can be applied upstream of the conversion from binary to mixed-radix,
while the readout process (errors from which may reintroduce repeats and prohibited motifs)
can be integrated into the decoding model as a step that follows the conversion from mixed-radix to DNA.

Our strategy uses several techniques borrowed from other areas of bioinformatics and information theory:
finite-state transducers, De Bruijn graphs, arithmetic coding, and synchronization codes.
Foremost among these are transducers \cite{MohriPereiraRiley2000,WikipediaTransducers}, automata-theoretic models of sequence transformation
which we use to represent in a uniform way the various steps of error-correction (e.g. via introduction of parity or watermark bits),
radix conversion, repeat-avoidant nucleotide encoding, and random sequencing error.
Standard algorithms for combining and decoding transducers can then be applied.
Transducers have been previously used in bioinformatics for
protein classification \cite{EskinEtAl2000},
phylogenetic indel models \cite{PatenEtAl2008,WestessonEtAlArxiv2012,WestessonEtAl2012},
and cancer informatics \cite{SchwarzEtAl2014}
To build the transducer that converts a mixed-radix sequence into a non-repeating DNA sequence,
we make use of De Bruijn graphs
and their relationship to finite-context automata (the probabilistic versions of which are known as order-$N$ Markov models).
De Bruijn graphs are widely used for genome assembly \cite{DeBruijn1946,PevznerEtAl2001,ZerbinoBirney2008,IqbalEtAl2012},
while order-$N$ Markov models are often employed for gene-finding \cite{BurgeKarlin1997,SalzbergEtAl1999}.
To build the transducer that converts a binary sequence into a mixed-radix sequence,
we borrow concepts from the arithmetic coding algorithm \cite{Rissanen1976,Mackay2003}.

\section{Methods}

\secref{Preliminaries} reviews preliminary concepts such as operations on weighted finite-state transducers,
De Bruijn graphs, patterns in DNA sequences, and the arithmetic coding algorithm.
\secref{DeBruijnTransducer} describes the repeat-avoiding transducer that accepts a mixed-radix sequence on its input
and generates a DNA sequence without homopolymers, short microsatellites or other local repeats.
\secref{DelayedMachine} reformulates this transducer so each state has ``knowledge'' of its past and future context,
facilitating error models that use this context to identify local duplications or context-sensitive substitutions.
\secref{Arithmetic} develops a transducer motivated by arithmetic coding
that converts a binary input sequence into a mixed-radix sequence suitable for subsequent encoding as DNA.
\secref{ErrorModel} describes an error-model transducer that mutates and deletes bases and reintroduces repeats.
\secref{Watermarks} describes a transducer that introduces markers and a watermark pilot sequence for synchronization.
\secref{CombinedCode} shows how all these transducers can be combined.

\subsection{Preliminary concepts}
\seclabel{Preliminaries}

\subsubsection{Weighted finite-state transducers}

Following \cite{MohriPereiraRiley2000}:
Assume a general semiring
$\semiring=(\srset,\srplus,\srtimes,\srplusid,\srtimesid)$
which for our purposes is typically the probability semiring
$(\Re,+,\times,0,1)$
or the tropical semiring
$(\Re_{+} \cup {\infty},\min,+,\infty,0)$.

A weighted finite-state transducer is defined as a tuple
$\trt = (\inalph,\outalph,\states,\transitions,\initstate,\finalstate)$
consisting of an input alphabet $\inalph$,
an output alphabet $\outalph$ (both alphabets being finite sets),
a finite set of states $\states$,
a finite set of transitions
$\transitions \subseteq \states \times (\maybe{\inalph}) \times (\maybe{\outalph}) \times \srset \times \states$,
an initial state $\initstate \in \states$
and a final state $\finalstate \in \states$.

The transducer $\trt$ can be thought of as an edge-labelled directed graph
where each state is a vertex
and each transition
$\trans = (\src[\trans],\inlab[\trans],\outlab[\trans],\weight[\trans],\dest[\trans]) \in \transitions$
is an edge from state $\src[\trans]$ to state $\dest[\trans]$
with input label $\inlab[\trans]$,
output label $\outlab[\trans]$
and weight $\weight[\trans]$.

A path in $\trt$ is a series of transitions that form a path in this graph.
The input sequence and output sequence for a path are the concatenation of (respectively)
the input and output labels of the transitions in the path.
The path weight is the $\srtimes$-product of the transition weights.
A successful path is one that starts in $\initstate$ and ends in $\finalstate$.
The transduction weight for a given input sequence $\inseq \in \inseqs$
and output sequence $\outseq \in \outseqs$
is the $\srplus$-sum of all successful paths
having $\inseq$ as the input sequence
and $\outseq$ as the output sequence.
Thus $\trt$ provides a mapping
$\tfunc{T}:(\kleene{\inalph} \times \kleene{\outalph}) \to \srset$
from sequence-pairs to weights.
We call this mapping $\tfunc{T}$ the transducer function.
For a given pair of sequences $(\inseq,\outseq)$
and a semiring wherein $\srplus$ and $\srtimes$ are amortized-constant resource operations,
it can be evaluated in time $\bigo(|\inseq| \cdot |\outseq| \cdot |\transitions|)$
and memory $\bigo(|\inseq| \cdot |\outseq| \cdot |\states|)$
by dynamic programming,
analogously to the Forward algorithm in the probabilistic semiring
or the Viterbi algorithm in the tropical semiring
\cite{Durbin98}.

A state $\state \in \states$ has past input context $\seqpast$ if every path from $\initstate$ to $\state$ has an input sequence with suffix $\seqpast$,
and future input context $\seqnext$ if every path from $\state$ to $\finalstate$ has an input sequence with prefix $\seqnext$.
The past output context and future output context of a state are defined similarly.

A ``waiting state'' is a state that has no outgoing transitions with empty input labels.
A ``waiting machine'' is a transducer where all the transitions with nonempty input labels
originate from waiting states.
Any transducer $\trt$ with a transducer function $\tfunc{T}$
can be transformed into an equivalent waiting machine
that has the same transducer function $\tfunc{T}$ and
at most $2|\states|$ states and $|\states|+|\transitions|$ transitions
\cite{WestessonEtAlArxiv2012}.

\subsubsection{Transducer composition}
\seclabel{TransducerComposition}

Given transducers
 $\trr = (\inalph, \somealph, \tpr{\states}, \tpr{\transitions}, \tpr{\initstate}, \tpr{\finalstate})$ and
 $\trs = (\somealph, \outalph, \tps{\states}, \tps{\transitions}, \tps{\initstate}, \tps{\finalstate})$
where $\trr$'s output alphabet is the same as $\trs$'s input alphabet,
we can readily find a composite transducer
 $\trt = \trr \transcomp \trs = (\inalph, \outalph, \tpt{\states}, \tpt{\transitions}, \tpt{\initstate}, \tpt{\finalstate})$
such that, if $\tfunc{R}$, $\tfunc{S}$ and $\tfunc{T}$ are the corresponding transducer functions,
then
\[
\forall \inseq \in \inseqs, \outseq \in \outseqs:
\quad
\tfunc{T}(\inseq,\outseq) = \srsum_{\someseq \in \someseqs} \tfunc{R}(\inseq,\someseq) \tfunc{S}(\someseq,\outseq)
\]
that is, $\trt$ models the feeding of $\trr$'s output into $\trs$'s input
(and this intermediate sequence is then summed out---i.e. marginalized, if we are in the probabilistic semiring).

Loosely speaking, we can construct $\trt$ using the following recipe:
\begin{itemize}
\item Each $\trt$-state corresponds to a pair of $\trr$- and $\trs$-states,
so $\tpt{\state} = (\tpr{\state}, \tps{\state})$
and $\tpt{\states} \subseteq \tpr{\states} \times \tps{\states}$.
\item The initial $\trt$-state $\tpt{\initstate}=(\tpr{\initstate},\tps{\initstate})$ pairs the initial states of $\trr$ and $\trs$.
\item The final $\trt$-state $\tpt{\finalstate}=(\tpr{\finalstate},\tps{\finalstate})$ pairs the final states of $\trr$ and $\trs$.
\item $\trt$-transitions
$\tpt{\trans} = ((\tpr{\src},\tps{\src}),\inlab,\outlab,\tpt{\weight},(\tpr{\dest},\tps{\dest}))$
can represent synchronized pairs of $\trr$-transitions
$\tpr{\trans} = (\tpr{\src},\inlab,\midlab,\tpr{\weight},\tpr{\dest})$
and $\trs$-transitions
$\tps{\trans} = (\tps{\src},\midlab,\outlab,\tps{\weight},\tps{\dest})$
which share the same intermediate label $\midlab$.
The composite transition weight $\tpt{\weight}$ is the product $\tpr{\weight} \srtimes \tps{\weight}$.
\item $\trt$-transitions can also represent transitions wherein only one of $\trr$ or $\trs$ changes state.
In these transitions, the intermediate label ($\midlab$) and either the input or output label ($\inlab$ or $\outlab$)
must be empty.
Some further synchronization may be required; for example, we can require that $\trs$ is a waiting machine,
and that $\trr$ can only change state when $\trs$ is in a waiting state \cite{WestessonEtAlArxiv2012}.
\end{itemize}

This algorithm can be made precise enough to automate;
more detailed workings are given elsewhere \cite{PereiraRiley1996,MohriPereiraRiley2000,Holmes2003,Holmes2007,WestessonEtAlArxiv2012,WestessonEtAl2012}.
The transducers yielded by a fully automated implementation can typically be aggressively optimized by hand
to minimize the number of transitions and/or states,
and hence,
the time and/or memory complexity of dynamic programming algorithms.

\subsubsection{Transducer concatenation}

Another operation on transducers that is useful in constructing DNA codes
is concatenation.
Suppose
 $\tra = (\inalph, \outalph, \tpa{\states}, \tpa{\transitions}, \tpa{\initstate}, \tpa{\finalstate})$ and
 $\trb = (\inalph, \outalph, \tpb{\states}, \tpb{\transitions}, \tpb{\initstate}, \tpb{\finalstate})$
are transducers with the same input and output alphabets and disjoint state spaces.
The concatenated transducer $\trt = \tra \transconcat \trb$ can be constructed by
taking the union of the state spaces and adding an $\epsilon$-labeled unit weight transition
from $\tpa{\finalstate}$ to $\tpb{\initstate}$.
This models feeding the first part of a sequence through $\tra$ and then feeding the second part through $\trb$.

\subsubsection{Transducer union}

The union of two transducers (whose state spaces are assumed disjoint)
is found by taking the union of their state spaces and transition graphs
and adding a new initial state with two unit-weight transitions, one to each of the initial states
of the two transducers being combined.

\subsubsection{Kleene closure}

The operation analogous to Kleene closure (generating all strings over a given alphabet)
is simply to add a unit-weight transition from the final state of the transducer back to the initial state.

\subsubsection{De Bruijn graphs}

Denote by $\somealph^\kmerlen$
the set of all possible $\kmerlen$-symbol strings
$\sym_1 \sym_2 \ldots \sym_\kmerlen$ over some alphabet $\somealph$.

The $\kmerlen$-dimensional De Bruijn graph over $\somealph$
has vertex set $\somealph^\kmerlen$
and a directed edge $u \to v$ between any two vertices
$u=\sym_1 \sym_2 \ldots \sym_\kmerlen$ and $v=\sym_2 \ldots \sym_\kmerlen \sym_{\kmerlen+1}$
that overlap by $\kmerlen-1$ symbols; this edge is labeled with symbol $\sym_{\kmerlen+1}$ (the last symbol of $v$).
Thus, each vertex has $|\somealph|$ incoming and $|\somealph|$ outgoing edges \cite{DeBruijn1946,PevznerEtAl2001}.
Denote this graph by $\debruijngraph = (\somealph^\kmerlen,\debruijnedges)$.

\subsubsection{DNA alphabets and repeats}

Let $\nucalph = \{ \mbox{A}, \mbox{C}, \mbox{G}, \mbox{T} \} $ be the nucleotide alphabet.
Let $\comp{x}$ denote the complement of a nucleotide symbol $x$,
and $\revcomp{\someseq}$ the reverse complement of a nucleotide sequence $\someseq$.
Let $\ntrans{x}$ denote the nucleotide related to $x$ by a transition substitution
(so $\ntrans{\nuca} = \nucg$, $\ntrans{\nucc}=\nuct$, etc.).

A direct tandem repeat of length $k$ is a nucleotide sequence followed by an exact copy of itself, $\someseq \someseq$, where the length of each copy is $\seqlen{\someseq}=k$
(note this includes repeated single nucleotides when $k=1$).
Similarly, a direct inverted repeat of length $k$ is a $k$-nucleotide sequence followed by its reverse complement, $\someseq \revcomp{\someseq}$.
A local inverted repeat of length $k$ and separation $l$ is a $k$-nucleotide sequence, followed an $l$-nucleotide sequence, followed by the reverse complement
of the first sequence; that is, $\someseq \otherseq \revcomp{\someseq}$, where $\seqlen{\someseq}=k$ and $\seqlen{\otherseq}=l$.

These repeats arise very commonly in naturally occurring DNA (e.g. see \cite{Ellegren2004}),
and are also hotspots for sequencing error \cite{LaehnemannEtAl2016}.

\subsubsection{Arithmetic coding}
\seclabel{ArithmeticCoding}

Arithmetic coding is a method for compressing a message $x_1 x_2 \ldots x_N$
given a probability model over messages whose predictive marginals for the next character in the sequence, $Q_n(k) \equiv P(x_{n+1}|x_1 \ldots x_n)$,
can be efficiently evaluated.
Suppose without loss of generality that the symbol alphabet (including the end-of-block character) is the set of integers $1 \ldots K$
and let $C_n(k) \equiv P(x_{n+1} \leq k | x_1 \ldots x_n)$ be the cumulative distributions for the predictive marginals.
Arithmetic coding uses the following ideas:
\begin{itemize}
\item The probability distribution over the first character divides the interval $[0,1)$ into $K$ subintervals.
  The $k$'th subinterval is $[C_1(k-1),C_1(k))$.
  \item Each subinterval is then further subdivided by the second character, then the third, and so on until the end of message.
    So, for example, the subinterval for the first character is $[C_1(x_1-1),C_1(x_1))$,
      for the second character $[C_1(x_1-1) + Q_1(x_1) C_2(x_2-1),C_1(x_1) + Q_1(x_1) C_2(x_2))$,
          and so on.
        \item If the interval before encoding the $n$'th character is $[A_n,B_n)$ then
          $[A_0,B_0) = [0,1)$,
              $A_{n+1} = (B_n-A_n) C_{n+1}(x_{n+1}-1)$,
              and $B_{n+1} = (B_n-A_n) C_{n+1}(x_{n+1})$.
            \item The upshot of all of this is that each message is associated with a finite subinterval of $[0,1)$
              and these subintervals are lexicographically ordered (i.e. sorted by the first symbol, then the second, then the third, and so on).
            \item Any finite-precision floating-point number also specifies a subinterval of $[0,1)$
              containing all floating-point numbers of greater precision that would be rounded to that number.
              For example, the decimal floating-point number $0.413$ specifies the subinterval $[0.4125,0.4135)$.
              \item To encode the message $x_1 x_2 \ldots x_N$,
                we simply need to transmit enough digits of a (binary) floating-point number such that its
                rounding interval is fully contained within the interval $[A_{N+1},B_{N+1})$.
                  The number of floating-point binary digits this requires should not exceed the Shannon entropy of the message,
                  $-\log_2 P(x_1 x_2 \ldots x_N)$, by more than one bit.
\end{itemize}

\subsection{A transducer for encoding signals as DNA without short repeats or reserved words}
\seclabel{DeBruijnTransducer}

In this section we contruct, in several steps, a transducer $\transcoder$
that accepts a mixed-radix sequence on the input
(that is, every input state either accepts binary symbols $\{ \bit{0}, \bit{1} \}$,
ternary symbols $\{ \trit{0}, \trit{1}, \trit{2} \}$ or
quaternary symbols $\{ \quat{0}, \quat{1}, \quat{2}, \quat{3} \}$)
and outputs a (uniquely decodable) nucleotide sequence that is free of
repeated nucleotides, short tandem repeats, or inverted repeats.
The input sequence may optionally be interleaved with special control digits
$\{ \controlsym{i}:\ 1 \leq i \leq \ncontrols \}$
which force the machine to output predictable, recognizable nucleotide sequences
that can be used to flag metadata, demarcate boundaries,
or to embed biologically functional motifs.

Start with $\debruijngraph$, the $\kmerlen$-dimensional De Bruijn graph over $\nucalph$.
Delete all nodes corresponding to sequences that are (or contain substrings which are)
direct tandem repeats of any length,
direct inverted repeats of length $\geq 2$,
or local inverted repeats of length $\geq \invreplen$ and separation $\geq 2$.
Denote the resulting graph by $\norepgraph = (\norepvertices,\norepedges)$.

Let $A \subset \norepvertices$ be a set of vertices to avoid,
let $D \in A$ be a target vertex,
and denote by $\prequels(D,A,N) \subset \norepvertices$
the set of all vertices
from which there exists a length-$N$ path to $D$
that does not pass through any of the vertices in $A$
(although the path is allowed to originate from one of those vertices).
Define $\stepsto(D,A)$ to be the smallest value of $N$ for which $\prequels(D,A,N) = \norepvertices$,
if such a value of $N$ exists; otherwise, let $\stepsto(D,A)$ be $\infty$.
We can find $\prequels(D,A,N)$ from $\prequels(D,A,N-1)$ by recursive backtracking,
and so determine in $k$ steps whether $\stepsto(D,A) \leq k$.

We now allocate $\ncontrols$ ``control words'': $\kmerlen$-mers
that will be reserved for marking metadata boundaries,
such as the start and end of messages.
In the transducer, these will be unreachable except by specially constructed paths of uniform length.
Specifically, we seek an indexed set of vertices
$\controlset = \{ \controlword_1, \controlword_2 \ldots \controlword_{\ncontrols} \}$
such that $\stepsto(\controlword_n,\controlset) \leq k$ for all $n$
and for some pre-specified value of $k$.
Our implementation finds this list $\controlset$ by brute-force recursive search
(using $k = 2\kmerlen$)
and further attempts to maximize the shortest Hamming distance between any two control words in $\controlset$.
Note that there is a ceiling to the number of control words that may be found
for any given value of $\kmerlen$ (and $\invreplen$),
though this ceiling grows rather rapidly with $\kmerlen$.
In practice we only need a few control words for most purposes.

Having designated some $\kmerlen$-mers as control words via analysis of $\norepgraph$,
we now construct a new graph $\controlgraph$
in which the control words are unreachable from the other words
except by paths that we construct.
Starting with graph $\norepgraph$, delete all incoming transitions to control words $\controlword_n \in \controlset$,
rendering them unreachable,
and prune the graph of any other nodes that become unreachable as a result.
Next, for every control word $\controlword_n \in \controlset$
and every path length $1 \leq k < \stepsto(\controlword_n,\controlset)$,
we create a new vertex set $\controlbridges{n,k}$
with a one-to-one correspondence to $\prequels(\controlword_n,\controlset,k)$.
We connect the newly-added vertices such that there is an edge from
$u \in \controlbridges{n,k}$ to $v \in \controlbridges{n,k-1}$
for every corresponding edge $(u',v') \in \norepedges$ between
$u' \in \prequels(\controlword_n,\controlset,k)$
and
$v' \in \prequels(\controlword_n,\controlset,k-1)$.
We also add edges from
$u \in \norepvertices$ to $v \in \controlbridges{n,\stepsto(\controlword_n,\controlset)}$
for the first steps in the paths to control words,
as well as edges from
$u \in \controlbridges{n,1}$ to $\controlword$
for the final steps.
In each case the newly-added edge is copied from a corresponding edge in $\norepedges$
and inherits the same edge label.

It can be useful to force the transducer to start with a particular control word $\initword$
and finish with another (or the same) control word $\finalword$.
To guarantee this we can add a chain of initial vertices and edges leading from source vertex to the initial control word,
$\initvertex{0} \to \initvertex{1} \to \initvertex{2} \to \ldots \to \initvertex{\kmerlen-1} \to \initword$,
with each transition labeled with consecutive symbols from the initial control word.
We also need to add a transition from the final control word to the final state.
(If the start and end control words are the same, then this vertex should be duplicated to prevent cycles.)

We have now constructed the graph $\controlgraph$ from which our transducer $\transcoder$
is derived via the following recipe:
\begin{itemize}
\item Every vertex in $\controlgraph$ is a state in $\transcoder$.
The initial and final vertices of $\controlgraph$ are the initial and final states of $\transcoder$.
\item Every edge in $\controlgraph$ is a unit-weight transition in $\transcoder$.
The output label of the transition is the label of the edge.
\item The input label of each transition is determined as follows:
\begin{itemize}
\item For states in $\norepvertices$ with two outgoing transitions to other states in $\norepvertices$, those transitions are input-labeled with the binary digits (bits) $\bit{0}$ and $\bit{1}$.
\item For states in $\norepvertices$ with three outgoing transitions to other states in $\norepvertices$, those transitions are input-labeled with the ternary digits (trits) $\trit{0}$, $\trit{1}$ and $\trit{2}$.
\item For states in $\norepvertices$ with four outgoing transitions to other states in $\norepvertices$, those transitions are input-labeled with the quaternary digits (quats) $\quat{0}$, $\quat{1}$, $\quat{2}$ and $\quat{3}$.
\item Transitions from states in $\norepvertices$ to states in $\controlbridges{n,\stepsto(\controlword_n,\controlset)}$,
which begin a path to the $n$'th control word $\controlword_n$,
are input-labeled with the special control digit $\controlsym{n}$.
\item All other transitions have input label $\emptystring$.
\end{itemize}
\end{itemize}

Thus, the input alphabet of $\transcoder$
is $\{ \bit{0}, \bit{1},
       \trit{0}, \trit{1}, \trit{2},
       \quat{0}, \quat{1}, \quat{2}, \quat{3},
       \controlsym{1} \ldots \controlsym{\ncontrols} \}$.

\figref{DNAStore} illustrates some of the code transducers that are generated by this procedure for the simplest case $\kmerlen=2$.
       
\subsubsection{Trading past context for future context}
\seclabel{DelayedMachine}

By virtue of derivation from the De Bruijn graph,
most of the states in the transducer $\transcoder$ of \secref{DeBruijnTransducer}
have $\kmerlen$ nucleotides of past output context.
We can transform this transducer into an equivalent one $\transdelay$ wherein most of the states have
$\kmerlen/2$ nucleotides of past output context
and $\kmerlen/2$ nucleotides of future output context.

This can be a convenient way to think about context, for the purpose of modeling errors in DNA replication and sequencing.
Common decoding and replication errors include local tandem and inverted duplications, as well as context-dependent substitutions.
These can occur on either strand of the DNA double helix and therefore (depending on the representation) may be best represented as depending
on future context as well as past context.

The transformation requires that the output sequence of all successful paths through the transducer
begin with a particular $\kmerlen$-mer and end with a particular $\kmerlen$-mer,
which can be ensured using the method described in \secref{DeBruijnTransducer}.

Let $\seqpastsyms{1}{\kmerlen} = \seqpast$ be the past output context for a given state $\state$.
All of the transitions $\trans$ into $\state$ have the same input label $\inlab[\trans] = \sympast_{\kmerlen}$,
which corresponds to the most recent nucleotide of output context for that state.
If, instead, we set $\inlab[\trans]$ to be $\sympast_{\kmerlen/2}$, the $(\kmerlen/2)$'th symbol of $\someseq$,
for all transitions into $\state$, then we have effectively delayed all output by $\kmerlen/2$ symbols.
We can now predict the next $\kmerlen/2$ symbols in the output sequence for the path from a state,
so we have traded $\kmerlen/2$ of past output context for $\kmerlen/2$ of future output context.

We need to add $\kmerlen/2$ extra padding states after (what was originally) the final state,
with a chain of transitions that flushes out the final $\kmerlen/2$ delayed output symbols.
Conversely, the first $\kmerlen/2$ transitions from the initial state
(into states with $\kmerlen/2$ or fewer nucleotides of output context)
will, after the transformation,
be null transitions (with both input and output labels equal to $\emptystring$),
so these transitions and the corresponding states can be deleted.

An analogous procedure can be used to transform a machine with $\kmerlen$ symbols of past input context
into a machine with $\kmerlen/2$ past input context and $\kmerlen/2$ future input context.

\subsection{A transducer that converts a binary sequence into a mixed-radix sequence of binary, ternary and quaternary digits}
\seclabel{Arithmetic}

We here describe a transducer $\transrad$ for converting the binary sequence to the mixed-radix sequence
that we use to encode it in DNA that is free of short repeats.
We can always convert a binary sequence to a mixed-radix sequence trivially by encoding
$\bit{0}$ and $\bit{1}$ as $0_R$ and $1_R$ in whatever radix $R$ is available;
that is, we simply never use the extra symbols in our alphabet $\{ \trit{2}, \quat{2}, \quat{3} \}$.
However, we can do better than this and pack some extra information into the additional symbols when they are available.
There are at least two issues we need to consider here: (i) efficient conversion between binary and other radices
(especially ternary, which is the trickiest) via block codes, while allowing for prematurely truncated blocks;
(ii) the fact that our repeat-avoiding DNA code is mixed-radix, that is, it has a radix that varies from position to position.

Considering first the issue of conversion from binary to ternary,
it is tempting to try something like the machine of \figref{BinaryToTrinary},
which maps binary $\bit{0}\bit{0}$ to ternary $\trit{0}$,
$\bit{0}\bit{1}$ to $\trit{1}$,
and $\bit{1}$ to $\trit{2}$;
that is, it sometimes encodes two bits as one trit.
In fact, it does this exactly half the time (assuming a uniform distribution over input bits)
so there are, on average, 1.5 input bits per output trit,
which is quite close to the theoretical maximum of $\log_2(3) \simeq 1.585$ bits per trit.
However, we have to be a little careful with this approach:
the machine of \figref{BinaryToTrinary} can get stuck in a state where it has queued up a zero input bit
and is waiting for the next input bit before it outputs a trit.
If the input ends at this point, or if the encoded signal includes a non-bit symbol
such as one of the reserved control symbols we allowed for in \secref{DeBruijnTransducer},
then we have to flush out that queued-up zero bit somehow.

A fragile workaround to this problem is to send an extra padding bit in such cases;
more robustly, we can introduce an end-of-block symbol `\$',
perhaps as part of a block code.
Of course, once we start adding additional symbols like end-of-block, the number of input bits we can encode per output trit (or output bit, or quat) has to fall,
since we have to reserve some codewords for transmitting these symbols.

The Huffman block code for binary encoding is well known, and generalizes readily to ternary \cite{BornholtEtAl2016}.
This brings us, however, to the second issue we face in radix conversion.
A ternary Huffman code is imperfectly suited to the coding of information of DNA wherein we are trying to avoid dinucleotide, trinucleotide or higher-order repeats;
nor is it well-suited to the situation where we are allowing repeats but avoiding certain reserved words.
The reason for the mismatch is that, in such situations, the number of available nucleotides at any given position may vary between 1, 2, 3 or 4.
Positions where only one nucleotide is legal cannot carry information; but if there are 2, 3 or 4 options,
then we need to think about encoding a binary, ternary, or quaternary digit, as appropriate.

For mixed-radix output, in the special case where the radices at each position are fully specified,
the relevant cousin of the Huffman code is the mixed-radix Huffman code, for the discovery of which optimal dynamic programming strategies have been presented \cite{ChuGill1992,GolinEtAl2009}.
However, taking this approach in our situation would require a different mixed-radix Huffman code for every possible combination of radices,
and so is not particularly convenient for the situation when the number of available coding nucleotides varies from site to site
(as in the repeat-avoiding code).

In this section, we present a strategy for developing compact automata (transducers) that convert short binary codewords into any mixed-radix encoding.
The code approaches optimality at long block lengths.
Unfortunately, the number of states used by the machine grows exponentially with the sequence length, so this asymptotic limit is hard to realize.
Nevertheless, the codes generated are not totally awful, and approach the asymptotic limit reasonably fast,
as shown in \secref{Results}.

We apply the concepts of arithmetic coding (\secref{ArithmeticCoding}) to encode input words as mixed-radix floating point numbers.
The set of input words to be encoded is the set of all length-$N$ binary strings,
plus the set of all length-$k$ binary strings ($0 \leq k < N$) that are followed by the `\$' character.
Thus, if $N=2$ then the input word set is $\{\$,0\$,1\$,00,01,10,11\}$.
There are $2^{N+1}-1$ words in the input word set.

We assume that a probability model as described above is defined over this input set;
typically we will use a simple model that assigns a small probability $\nu$ to `\$' at any position,
independently of previous context,
and splits the remaining probability evenly between $0$ and $1$.
Each input word $w$ is then associated with an interval $[A_w,B_w)$ of size $P(w) = B_w-A_w$.

Since we are working with small fixed-length input codewords and finite-state machines, we will drop the lexicographical ordering of codeword intervals
which arithmetic coding uses to implement online compression and decompression for arbitrary-length messages.
Instead, we sort the input words $w$ by probability $P(w)$, which will tend to lead to rarer words sharing similar encoded prefixes
(c.f. Huffmann coding).
We also apply several optimizations (dynamically adjusting the probability distribution to improve performance on rare input words,
merging identical states in the transducer, and pruning unnecessary states) that help offset the inefficiency of using a short block code
(as compared to arithmetic coding where the block code length is effectively the length of the entire message).

\begin{algorithm}
  \KwData{$W$, the list of input words sorted by decreasing probability}
  \KwData{$A_w,B_w,P(w)$, the bounds and size of the probability interval for each input word}
  \KwResult{A finite-state transducer that converts an input word into a mixed-radix sequence with all possible combinations of binary, ternary and quaternary radix at every available position}

  First construct the prefix tree of the input words, assign a state to every node and an input-labeled transition to every correspondingly labeled edge.
The root of the prefix tree serves as the initial and final state
\;
\For{$w \in W$}{
  Let $l$ be the leaf $l$ of the prefix tree associated with $w$
  \;
  The {\em input interval} is $[A_w,B_w)$
    \;
    Let $m = A_w + \frac{1}{2}(B_w - A_w)$ be its midpoint
    \;
    Let $S$ be the list of states still to be processed
    \;
    Let $F$ be the set of final states
    \;
    Let $O_s$ denote the {\em output prefix} of state $s$
    \;
    Let $[D_s,E_s)$ denote the {\em output interval}
      \;
      {\em (The output prefix is the sequence of symbols that must be emitted to reach $s$.
        The output interval is the subinterval of $[0,1)$ that has been encoded by the output prefix)}
      \;
      Initialize $S \leftarrow \{ l \}$, $O_l \leftarrow \emptystring$, $D_l \leftarrow 0$, $E_l \leftarrow 1$
          \;
          \While{$S \neq \emptyset$}{
            Let $s$ be the first state in $S$
            \;
            Delete $s$ from $S$
            \;
            \eIf{$D_s \geq A_w$ {\bf and} $E_s \leq B_w$}{
              Add $s$ to $F$.}{
              \For{$R \leftarrow 2$ \KwTo $4$}{
                Find the integer $r$ such that if $d(r)=D_s+r(E_s-D_s)/R$ then $d(r) \leq m < d(r+1)$
                \;
                Create a new state $t$
                \;
                Add a transition from $s$ to $t$ with output label $r_R$
                \;
                Set $D_t \leftarrow d(r)$, $E_t \leftarrow d(r+1)$ and $O_t \leftarrow O_s \cdot r_R$
                \;
                Add $t$ to $S$.
              }
            }
          }
                {\em (Having generated all mixed-radix encodings for $w$,
                  we can now shrink its input interval
                  to just enclose the output intervals used by the encodings.
                  Its upper bound drops from $B_w$ to $E_{\max}$,
                  increasing the space available for the remaining input words by a factor of $\alpha$)}
          \;
          Let $E_{\max} = \max_{s \in F} E_s$ and $\alpha = (1 - E_{\max}) / (1 - B_w)$
          \;
          \For{$x \in W$, $A_x > A_w$}{
            $A_x \leftarrow \alpha(A_x - B_w) + E_{\max}$
            \;
            $B_x \leftarrow \alpha(B_x - B_w) + E_{\max}$
          }
        }
    If any states have a unique output prefix $O_s$, then remove all their descendants
        \;
        Merge all states whose sets of possible output sequences are identical
        \;
        Form the Kleene closure of the constructed transducer.
\caption{
  \alglabel{MixedRadixTransducer}
  Algorithm to generate a transducer that converts from binary to a mixed-radix sequence
  (\secref{Arithmetic}).
}
\end{algorithm}

Pseudocode to generate the transducer is shown in \algref{MixedRadixTransducer}.
The essence of the approach is to build a tree of the digits of all possible radices that the arithmetic coding approach
would use to encode each position, with some optimizations (as noted) to merge states and adjust output intervals.
\figref{Arithmetic} shows a transducer that was built by this algorithm
for 2-digit codewords with $\nu=1/100$.

\subsection{A model of substitutions, local tandem and inverted duplications, and other indels}
\seclabel{ErrorModel}

In this section we describe a statistical error model,
implemented as a transducer $\transerror$,
that includes
tandem duplications (ACG$\to$ACG\underline{ACG}),
forward inverse duplications (ACG$\to$ACG\underline{CGT}),
reverse inverse duplications (ACG$\to$\underline{CGT}ACG),
and point substitutions (ACG$\to$A\underline{T}G).
The duplications can be imperfect: they can include substitutions
(e.g. ACG$\to$ACG\underline{A\underline{T}G}).
We also describe how to account for partial observation of the sequence
(at least at the level of reconstructing individual reads;
the broader problem of reassembling a data file from fragments is not addressed here,
beyond the general recipe for demarcating metadata with control words
that was given in \secref{DeBruijnTransducer}, which can be used to mark up shorter blocks
with their locations in the file).

The full working is rather detailed, but the basic idea is very simple.
We use the transducer state space to build up a context of past nucleotides
(i.e. the last few nucleotides it has most recently seen on the input)
and future nucleotides
(i.e. the next few nucleotides it is prepared to receive on the input, in a given state).
The higher-order structure of the transducer is determined by how much past and future context it has built up
(in the early stages),
and how much future context it has yet to work through
(in the later stages).
The local structure of the transducer involves states for
substitutions, deletions
and insertions (which are actually duplications that copy the local context).

The transducer as constructed can only handle sequences whose length is at least $\kmerlen$.
Shorter sequences will have no successful path through the machine.
This is not envisaged to be a problem for decoding DNA-stored data
since reads of length $<\kmerlen$ will contain very little data,
probably insufficient for assembly and lacking in metadata
since $\kmerlen$ is the number of nucleotides required to encode a control signal.
Nevertheless, the transducer can readily be adapted for such edge cases
by adding extra blocks to the higher-order structure (see \subfigref{PartialObservation}{a}).

The error-model transducer is defined over the probabilistic semiring,
has input alphabet $\outalph$,
output alphabet $\outalph$,
past \& future context $\contextlen=\kmerlen/2$,
and the following state space:
\begin{itemize}
\item There is a state $\symstate{a}(\emptystring,\emptystring)$ which is the initial state
\item For every $k:1 \leq k < \contextlen$ and every $k$-mer $\seqnext \in \outalph^k$
  there is a state $\symstate{b}(\emptystring,\seqnext)$
\item For every $k:1 \leq k < \contextlen$, every $k$-mer $\seqpast \in \outalph^k$ and every $\contextlen$-mer $\seqnext \in \outalph^\contextlen$
  there are
  \begin{itemize}
  \item two states $\{ \symstate{c}(\seqpast,\seqnext),\ \delstate{c}(\seqpast,\seqnext) \}$
  \item $\contextlen$ states $\{ \revstate{i}{c}(\seqpast,\seqnext,i):\ 1 \leq i \leq \contextlen \}$
  \item $2k$ states $\{ \tanstate{c}{i}(\seqpast,\seqnext),\ \fwdstate{c}{i}(\seqpast,\seqnext): 1 \leq i \leq k \}$
  \end{itemize}
\item For every pair of $\contextlen$-mers $\seqpast, \seqnext \in \outalph^\contextlen$
  there are
  \begin{itemize}
  \item two states $\{ \symstate{d}(\seqpast,\seqnext),\ \delstate{d}(\seqpast,\seqnext) \}$
  \item $3\contextlen$ states $\{ \tanstate{d}{i}(\seqpast,\seqnext),\ \fwdstate{d}{i}(\seqpast,\seqnext),\ \revstate{d}{i}(\seqpast,\seqnext):\ 1 \leq i \leq \contextlen \}$
  \end{itemize}
\item For every $k:1 \leq k < \contextlen$, every $\contextlen$-mer $\seqpast \in \outalph^\contextlen$ and every $k$-mer $\seqnext \in \outalph^k$
  there are
  \begin{itemize}
  \item two states $\{ \symstate{e}(\seqpast,\seqnext),\ \delstate{e}(\seqpast,\seqnext) \}$
  \item $2\contextlen$ states $\{ \tanstate{e}{i}(\seqpast,\seqnext),\ \fwdstate{e}{i}(\seqpast,\seqnext):\ 1 \leq i \leq \contextlen \}$
  \item $k$ states $\{ \revstate{e}{i}(\seqpast,\seqnext):\ 1 \leq i \leq k \}$
  \end{itemize}
\item There is a state $\symstate{f}(\emptystring,\emptystring)$ which is the final state
\end{itemize}
These states have the following significance (see \figref{PartialObservation}):
\begin{itemize}
  \item $\symstate{b}$ states load input symbols into the future context queue
  \item $\symstate{c}$ states have fully loaded future context queues. They continue to load input symbols, but also start emitting output symbols and shifting input symbols to the past context queue
  \item $\symstate{d}$ states have fully loaded past and future context queues. They load input symbols, shift input symbols from future to past context queues, drop input symbols off the back of the past context queue, and emit output symbols
  \item $\symstate{e}$ states have fully loaded past context queues, but emptying future context queues. No future input symbols are loaded at this point. They shift input symbols from future to past context queues, drop input symbols off the back of the past context queue, and emit output symbols
  \item \subfigref{PartialObservation}{a} shows an additional block of $\symstate{g}$-states that would be required for the transducer to handle sequences of length $<2\contextlen$, which never make it to block \#d since they are too short to fully load the past and future context queues. Since these sequences probably contain too little information to be useful (especially if we are using $2\contextlen$-nucleotide words to mark up metadata), we have omitted them from the formal description.
  \item $\delstate{\block}$ states are used for deletions ($\block \in \{ c, d, e \}$)
  \item $\tanstate{\block}{i}$ states are used for tandem duplications ($1 \leq i \leq \contextlen$)
  \item $\fwdstate{\block}{i}$ states are used for forward inverse duplications
  \item $\revstate{\block}{i}$ states are used for reverse inverse duplications
  \item Each state is indexed with its past context $\seqpast$, its future context $\seqnext$ and (for duplication states) the remaining duplication length $i$
\end{itemize}

The transitions $(p, \inlab, \outlab, \weight, q)$,
that involve $S$-states,
so $p = \symstate{\srcblock}(\ldots)$ and $q = \symstate{\destblock}(\ldots)$,
are shown in the table.
In all cases
$1 < k \leq \contextlen$ measures a partial context length,
$\seqpastsyms{1}{\contextlen} \in \outalph$ represent past input context symbols,
$\seqnextsyms{1}{\contextlen} \in \outalph$ represent future input context symbols and
$\outsym \in \outalph$ represents an output symbol.
\[
\begin{array}{lllll}
\src & \inlab & \outlab & \weight & \dest \\
\hline
\symstate{a}(\emptystring,\ \emptystring) & \symnext_1 & \emptystring & 1 & \symstate{b}(\emptystring,\ \symnext_1) \\
\symstate{b}(\emptystring,\ \seqnextsyms{1}{k-1}) & \symnext_k & \emptystring & 1 & \symstate{b}(\emptystring,\ \seqnextsyms{1}{k}) \\
\symstate{b}(\emptystring,\ \symnext_1\seqnextsyms{2}{\contextlen}) & \symnext_{\contextlen+1} & \outsym & \pcont(0,\contextlen) \cdot \psubnext{1} & \symstate{c}(\symnext_1,\ \seqnextsyms{2}{\contextlen}\symnext_{\contextlen+1}) \\
\symstate{c}(\seqpastsyms{1}{k-1},\ \symnext_1\seqnextsyms{2}{\contextlen}) & \symnext_{\contextlen+1} & \outsym & \pcont(k-1,\contextlen) \cdot \psubnext{1} & \symstate{c}(\seqpastsyms{1}{k-1}\symnext_1,\ \seqnextsyms{2}{\contextlen}\symnext_{\contextlen+1}) \\
\symstate{c}(\seqpastsyms{1}{\contextlen-1},\ \symnext_1\seqnextsyms{2}{\contextlen}) & \symnext_{\contextlen+1} & \outsym & \pcont(\contextlen-1,\contextlen) \cdot \psubnext{1} & \symstate{d}(\seqpastsyms{1}{\contextlen-1}\symnext_1,\ \seqnextsyms{2}{\contextlen}\symnext_{\contextlen+1}) \\
\symstate{d}(\sympast_1\seqpastsyms{2}{\contextlen},\ \symnext_1\seqnextsyms{2}{\contextlen}) & \symnext_{\contextlen+1} & \outsym & \pcont(\contextlen,\contextlen) \cdot \psubnext{1} & \symstate{d}(\seqpastsyms{2}{\contextlen}\symnext_1,\ \seqnextsyms{2}{\contextlen}\symnext_{\contextlen+1}) \\
\symstate{d}(\sympast_1\seqpastsyms{2}{\contextlen},\ \symnext_1\seqnextsyms{2}{\contextlen}) & \emptystring & \outsym & \pcont(\contextlen,\contextlen) \cdot \psubnext{1} & \symstate{e}(\seqpastsyms{2}{\contextlen}\symnext_1,\ \seqnextsyms{2}{\contextlen}\symnext_{\contextlen+1}) \\
\symstate{e}(\sympast_1\seqpastsyms{2}{\contextlen},\ \symnext_1\seqnextsyms{2}{k}) & \emptystring & \outsym & \pcont(\contextlen,k) \cdot \psubnext{1} & \symstate{e}(\seqpastsyms{2}{\contextlen}\symnext_1,\ \seqnextsyms{2}{k-1}) \\
\symstate{e}(\seqpastsyms{1}{\contextlen},\ \symnext_1) & \emptystring & \outsym & \pcont(\contextlen,1) \cdot \psubnext{1} & \symstate{f}(\emptystring,\ \emptystring) \\
\hline
\end{array}
\]

The other transitions can be deduced using the following rules:
\[
\begin{array}{llllll}
  \multicolumn{6}{l}{
    \text{For every state of the form...}
  } \\
\hline
\symstate{\block}(\seqpast,\seqnext) & & & & & \text{with $\seqpast = \seqpastsyms{1}{j},\ 0 \leq j \leq \contextlen$} \\
& & & & & \text{and $\seqnext = \seqnextsyms{1}{k},\ 0 \leq k \leq \contextlen$} \\
\hline
\\
\multicolumn{6}{l}{
  \text{...there are transitions of the form...}
} \\
\src & \inlab & \outlab & \weight & \dest \\
\hline
\symstate{\block}(\seqpast,\seqnext) & \emptystring & \emptystring & \ptandup \lendist(i) & \tanstate{\block}{i}(\seqpast,\seqnext) & \forall 1 \leq i \leq j \\
\symstate{\block}(\seqpast,\seqnext) & \emptystring & \emptystring & \pfwddup             & \fwdstate{\block}{1}(\seqpast,\seqnext) & \\
\symstate{\block}(\seqpast,\seqnext) & \emptystring & \emptystring & \prevdup \lendist(i) & \revstate{\block}{i}(\seqpast,\seqnext) & \forall 1 \leq i \leq k \\
\tanstate{\block}{i}(\seqpast,\seqnext) & \emptystring & \outsym & \psubpast{j+1-i} & \tanstate{\block}{i-1}(\seqpast,\seqnext) & \forall 1 < i \leq j \\
\tanstate{\block}{1}(\seqpast,\seqnext) & \emptystring & \outsym & \psubpast{j} & \symstate{\block}(\seqpast,\seqnext) & \\
\fwdstate{\block}{i}(\seqpast,\seqnext) & \emptystring & \outsym & \psubcomppast{j+1-i} & \fwdstate{\block}{i+1}(\seqpast,\seqnext) & \forall 1 \leq i < j \\
\fwdstate{\block}{1}(\seqpast,\seqnext) & \emptystring & \outsym & \lendist(i) \psubcomppast{j+1-i} & \symstate{\block}(\seqpast,\seqnext) & \forall 1 \leq i \leq j \\
\revstate{\block}{i}(\seqpast,\seqnext) & \emptystring & \outsym & \psubcompnext{i} & \revstate{\block}{i-1}(\seqpast,\seqnext) & \forall 1 < i \leq k \\
\revstate{\block}{1}(\seqpast,\seqnext) & \emptystring & \outsym & \psubcompnext{1} & \symstate{\block}(\seqpast,\seqnext) & \\
\hline
\\
\multicolumn{6}{l}{
    \text{For every transition of the form...}
  } \\
\src & \inlab & \outlab & \weight & \dest \\
\hline
\symstate{\srcblock}(\seqpast,\seqnext) & \sym & \outsym & \pcont(\ldots) \psub & \symstate{\destblock}(\seqpast',\seqnext') & \text{with $\seqpast,\seqpast',\seqnext,\seqnext' \in \kleene{\outalph}$} \\
  & & & & & \text{and $\sym,\outsym \in \outalph$} \\
\hline
\\
  \multicolumn{6}{l}{
    \text{...there are also transitions of the form...}
  } \\
\src & \inlab & \outlab & \weight & \dest \\
\hline
\symstate{\srcblock}(\seqpast,\seqnext) & \sym & \emptystring & \pdelopen & \delstate{\destblock}(\seqpast',\seqnext') & \text{if $\destblock \neq f$} \\
\symstate{\srcblock}(\seqpast,\seqnext) & \sym & \emptystring & \pdelopen & \symstate{f}(\emptystring,\emptystring) & \text{if $\destblock = f$} \\
\delstate{\srcblock}(\seqpast,\seqnext) & \sym & \emptystring & \pdelext & \delstate{\destblock}(\seqpast',\seqnext') & \text{if $\srcblock \neq b$} \\
\delstate{\srcblock}(\seqpast,\seqnext) & \emptystring & \emptystring & \pdelend & \symstate{\srcblock}(\seqpast',\seqnext') & \text{if $\srcblock \neq b$} \\
\hline
\end{array}
\]

The probability parameters are
$\pdelopen$ to open a deletion,
$\pdelext$ to extend a deletion,
$\pdelend$ to end a deletion,
$\ptandup$ for a tandem duplication,
$\pfwddup$ for a forward inverse duplication,
$\prevdup$ for a reverse inverse duplication,
$\pnogap$ for no gap,
$\ptransition$ for a transition substitution (A$\leftrightarrow$G or C$\leftrightarrow$T),
$\ptransversion$ for a transversion substitution (A$\leftrightarrow$C, A$\leftrightarrow$T, C$\leftrightarrow$G, G$\leftrightarrow$T),
$\pmatch$ for no substitution, and
$\plen{k}$ for the probability that a duplication has length $k$.
The constraints on these parameters are
$\pdelopen+\ptandup+\pfwddup+\prevdup+\pnogap=1$,
$\pdelext+\pdelend=1$,
$\ptransition+\ptransversion+\pmatch=1$,
and
$\sum_{k=1}^{\kmerlen} \plen{k} = 1$.

The substitution matrix is $\submat(\sym,\outsym)$, defined to be $\pmatch$ if $x=y$ (no substitution), $\ptransition$ if $x = \ntrans{y}$ (transition)
and $\ptransversion/2$ otherwise (transversion).

The context-adjusted probability of not opening a gap at a site with $j$ nucleotides of past context and $k$ nucleotides of future context is
\[
\pcont(j,k) = \pnogap + (\ptandup + \pfwddup) \sum_{i=j+1}^{\kmerlen} \lendist(i) + \prevdup \sum_{i=k+1}^{\kmerlen} \lendist(i)
\]

The probability parameters can be estimated directly from trusted alignments using the Baum-Welch algorithm \cite{Durbin98}.

To keep our implementation simple we used a basic error model.
It would be straightforward to give the error model a richer parameterization;
for example, allowing more free parameters in the substitution matrix,
or allowing the mutation parameters to depend on the adjacent context.

The transducer $\transerror$
described in this section models errors but still assumes observation of the full-length sequence.
By composing it with a 3-state transducer $L \to M \to R$ whose initial and final states ($L,R$) erase their input
and whose middle state ($M$) echoes its input unmodified,
we can model both errors and partial observation.

\subsection{Codes to correct insertion, deletion and substitution errors}
\seclabel{Watermarks}

In \secref{Arithmetic} we described a code that attempted to pack as much encoded information as possible into the mixed-radix sequence.
However, we can also use the extra information in the ternary and quaternary bits for error-correction purposes:
either to store parity bits, or to store a watermark.

The watermark was introduced by Davey and MacKay as an ``inner code'' to correct for insertion and deletion errors
in combination with an ``outer code'', such as a low-density parity check (LDPC) code, to correct substitution errors \cite{DaveyMackay2000,DaveyMackay2001}.
The watermark is a pilot sequence, known to encoder and decoder, which is XOR'd with a sparsified transform of the input signal
(so that some or most of the watermark is transmitted unmodified).
Deletions may corrupt a few bits of the input signal, but the extent of the deletion can be identified with reference to the watermark.
Watermark codes stand in contrast to ``marker'' codes, another strategy for insertion-deletion correction which explicitly flags the boundaries of blocks \cite{RatzerMackay2000}.

Using transducers we can readily introduce both watermark- and marker-style synchronization in several ways:
(i) as a marker approach that uses the ``reserved words'' developed in \secref{DeBruijnTransducer},
we can construct an $M$-state cyclic transducer that introduces a reserved word into the output DNA after every $M$ input bits;
(ii) as a watermark approach that uses the spare information in the higher-radix digits, we can give each of the $M$ states a unique pseudorandom signature for radix conversion,
so that every (binary) input digits is mapped uniquely to a single (binary, ternary or quaternary) output digit,
with this mapping performed in a pseudorandom way by each of the $M$ states;
(iii) for additional synchronization, we can use an approach closer to Davey and MacKay's original watermark technique
and have the transducer interleave input bits with watermark bits at some ratio.

When using synchronization codes, we first encode the message using an LDPC code,
which we then apply to the decoded sequence at the other end.
The block length of the LDPC code can be matched to the block length of the synchronization code
(i.e. the length of the watermark, or the spacing between markers)
to facilitate decoding of message fragments obtained via DNA sequencing.

\subsection{Combining the component transducers}
\seclabel{CombinedCode}

We have described several component transducers so far,
including machines
$\transrad$, for converting a sequence of binary symbols
into a mixed-radix sequence (\secref{Arithmetic});
$\transcoder$, for converting the mixed-radix sequence into nonrepeating DNA
(\secref{DeBruijnTransducer});
and $\transerror$ for modeling sequencing errors during decoding
(\secref{ErrorModel}).
Other relevant machines are depicted in the figures,
such as $\hammingsevenfour$ for implementing the Hamming(7,4) error-correction code
(\subfigref{HammingTransducer}{b}).

These machines can be combined by the process of transducer composition (\secref{TransducerComposition}) into a single integrated automaton suitable for coding.
For example, $\hammingsevenfour \transcomp \transrad \transcomp \transcoder \transcomp \transerror$
implements all four steps described in the previous paragraph.

The transducer composition procedure implicitly resolves the mixed-radix conversion.
Specifically, $\transrad$ converts a binary input into all possible mixed-radix sequences,
but $\transcoder$ will only accept one series of radices (the radix at each point in the sequence depending
on the most recent $\kmerlen$ nucleotides that have been emitted previously, and thus, the number of nucleotides available
at the next variable position without generating a tandem or inverted repeat).
Thus, only one of the possible output sequences of $\transrad$ is valid.
The selection of this valid output happens in the transducer composition $\transrad \transcomp \transcoder$
when we match transitions from $\transrad$ with those of $\transcoder$;
only transitions with compatible radices are used by this procedure.

The matching of transitions and states that occurs during transducer composition
implcitly discards large parts of the composite state space that are unreachable.
We can use this to particular advantage when combining $\transerror$ with $\transdelay$
(the version of $\transcoder$ described in \secref{DelayedMachine} that has both past and future context).
States in $\transerror$ will only mesh with states in $\transdelay$ that have compatible contexts,
so we can efficiently throw out many of the possible combinations of states.

While some error-correcting codes, such as short Hamming codes, can be conveniently represented
as state machines with exhaustively enumerable state spaces,
more sophisticated codes---such as LDPC codes---generally have too many possibilities to enumerate efficiently.
Instead, LDPC codes are generally decoded using sum-product belief propagation on cyclic graphical models \cite{Mackay1997,FreyMackay98}.
Following Davey and MacKay \cite{DaveyMackay2000},
we have here implemented decoding as a two-stage process wherein the HMM corrects indel errors
and the LDPC code corrects any remaining substitution errors.
Unlike Davey and MacKay, we do not currently make use of ``soft'' information from the
first decoding stage.
Since LDPC codes and HMMs can both be represented as factor graphs
of (respectively) trellis and cyclic topology \cite{Kschischang2006},
these could in principle offer a common likelihood framework in which uncertainty from the HMM decoding
could inform the second stage of LDPC bit-correction.

\section{Results}
\seclabel{Results}

We have developed open-source computer programs implementing the algorithms in this paper.
MIXRADAR implements the MIXed-RADix ARithmetic code of \secref{Arithmetic}.
DNASTORE implements the non-repeating DNA code of \secref{DeBruijnTransducer},
a subset of the error-correcting code of \secref{CombinedCode},
and the transducer composition algorithm from \secref{TransducerComposition}.
WATMARK generates transducers that implement the watermark synchronization of \secref{Watermarks}.

All these programs are available at \url{https://github.com/ihh/dnastore}
along with auxiliary codes (for example the Hamming(7,4) code of \subfigref{HammingTransducer}{b})
and scripts to generate those codes.

\subsection{MIXRADAR}
\seclabel{MIXRADAR}

\tabref{MIXRADAR.codes} shows statistics for codes generated with MIXRADAR.
It is straightforward to achieve (near-)ideal performance in converting bits to quats,
but to approach ideal performance in bit-to-trit conversion requires a block code.
Finding the optimal block length is equivalent to finding
positive integers $m,n$ for which $3^m$ is only a little more than $2^n$;
the number of bits/trit is then $n/m$.
For example, good values of $(m,n)$ are $(7,11)$
(since $3^7 \simeq 2^{11} \times 1.07$)
and $(12,19)$ (since $3^{12} \simeq 2^{19} \times 1.01$).
These offer bit/trit ratios of $11/7 \simeq 1.571$ and $19/12 \simeq 1.583$, respectively
(compared to the theoretical maximum $\log_2(3) \simeq 1.585$ bits/trit).
However, implementing codewords of 11 or 19 bits in length using a state machine
would require millions or billions of states
to encode all the possible mixed-radix outputs.
The best we can achieve at lower codeword lengths is $(m,n) = (4,6)$ (where $3^4 \simeq 2^6 \times 1.27$).
This codeword length (6 bits) gives a performance that is near-ideal for radix-2 output (1 bit/base) and radix-4 output (2 bits/base)
and decent for radix-3 output (1.5 bits/base), at 2125 states and 6375 transitions (see \tabref{MIXRADAR.codes}).

%



\subsection{DNASTORE}
\seclabel{DNASTORE}

In the absence of any other specified coding scheme,
DNASTORE uses the naive binary-to-ternary conversion of
\figref{BinaryToTrinary} and adds a zero padding bit when the machine needs to be flushed,
in contrast to the MIXRADAR block codes that do not require padding bits.
This can cause spurious single-bit insertion errors at the end of the message, and more serious problems when the
machine is composed with other coding schemes.
To circumvent such problems, and for general flexibility,
DNASTORE can be supplied with alternative coding machines described in a compact JSON format
and can compose multiple machines using the algorithm outlined in \secref{TransducerComposition}.
This allows the straightforward incorporation of composite state machines incorporating other stages
(e.g. more sophisticated radix conversion, or error correcting codes).

\tabref{DNASTORE.codes} shows statistics for codes generated with DNASTORE
using the algorithm of \secref{DeBruijnTransducer}.
For most of the simulations and composite codes in this paper, we have used a short code of $\kmerlen=4$ nucleotides,
which avoids single-base and dinucleotide repeats,
but not trinucleotide (or longer) repeats.

\subsection{Error correction}
\seclabel{Viterbi}

DNASTORE implements error correction using a simplified version of the transducer $\transerror$
of \secref{ErrorModel}, modeling only tandem duplications and not inverted duplications.
It can also estimate the parameters of the error model from training data
(in the form of alignments of sequenced DNA to known reference copies of that DNA).

To test the performance of these codes, we tried encoding and decoding a uniformly random sequence of 8,192 bits
while simulating the controlled corruption of the sequenced DNA by various types of random mutation:
point substitutions, non-overlapping deletions of up to 4 bases, and tandem duplications of up to 4 bases.
We explored several different composite codes, with components including
the length-4 and length-8 codes from \tabref{DNASTORE.codes};
the mixed-radix block code of 2-bit codewords from \tabref{MIXRADAR.codes};
the Hamming(7,4) error-correction code of \subfigref{HammingTransducer}{b};
the synchronization codes of \secref{Watermarks};
and an outer LDPC(16384,8192) code implemented using Neal's LDPC-codes
software available from \url{https://github.com/radfordneal/LDPC-codes}.
We varied substitution, deletion and insertion probabilities,
using the Baum-Welch algorithm to estimate the probability parameters from data,
repeating the experiment 20 times at each setting,
and measuring the Levenshtein edit distance (normalized by the number of input bits) between the original and recovered bit-sequences.

The results of these computational experiments are summarized in \figrange{DupPlot}{WatPlot}.
\figref{DupPlot} shows the decoder accuracy in the presence of simulated tandem duplication events.
Since the code (by construction) contains no tandem duplications of $\kmerlen$ or fewer bases,
the decoder (which models duplications as special types of error)
should be able to recover from these perfectly, and this is indeed the case---as long
as the maximum duplicated sequence length is less than the context length of the error decoder
(lower curve of \figref{DupPlot}).
When the duplication size exceeds the error model's context length,
the error model is unable to recognize them and duplications then significantly affect decoder accuracy
(upper curve of \figref{DupPlot}).

\figref{SubPlot} shows the performance of various codes in the presence of single-nucleotide substitution errors
scattered randomly throughout the sequence.
At very low substitution rates, the basic DNASTORE code (without any additional error-correction)
is still able to correct some errors, probably because those substitution errors introduce illegal sequences.
Composing a MIXRADAR code in front of the DNASTORE further improves the mitigation of substitution errors,
but both DNASTORE and MIXRADAR+DNASTORE codes quickly start to exhibit decoding errors as the substitution rate rises.
This can be contrasted with two explicitly error-correcting codes, based respectively on the Hamming(7,4) code
(which due to the low number of states can be represented as a state machine and incorporated directly into the
Viterbi decoder)
and a watermark code combined with an LDPC code
(the LDPC encoding and decoding take place independently of the main state machine).
These codes are signficantly more robust to substitution errors, maintaining a zero decoded error rate
up to around 0.2 substitutions per nucleotide.
The Hamming code performs slightly better than the LDPC code at the highest substitution rates,
which may be due to a combination of reasons:
(i) the Hamming code is implemented as a transducer (so error correction can be performed
as part of Viterbi decoding, and is therefore exact),
(ii) this simulation did not include any
``burst'' style errors, which would impact Hamming code performance more than LDPC,
(iii) the LDPC code starts to deteriorate significantly when it misses block boundaries
(and, again, the two-stage decoding protocol we used does not allow the LDPC decoder to influence the
Viterbi decoding, so any block boundary errors that the Viterbi decoder makes are final).

In \figref{DelPlot}, we investigated the same codes under simulated deletions.
Several points are notable from this plot.
First, deletions have a much bigger impact than substitutions on all codes
(each simulated deletion event removes from one to four bases, so the average number of deleted bases
is 2.5 times the x-axis labels on this plot; still, even accounting for this, the impact of deletions
is worse than the substitutions of \figref{SubPlot} and more comparable to the duplications of \figref{DupPlot}).
Second, unlike with substitutions, the Hamming code accuracy deteriorates even at low error rates---which
is to be expected, since deletions introduce burst errors that Hamming cannot correct.
Third, the watermark-based code does work as advertised to correct errors with 100\% accuracy---but
only at low deletion rates.
When the deletion rate exceeds 0.02 (meaning, with our average deletion size of 2.5 bases, that roughly
5\% of the bases are deleted) then the error rate starts creeping up.
And fourth, for all the codes tried, when the deletion rate exceeds around 0.13 (so roughly 1/3 of bases are deleted)
no signal is recoverable at all.

To further explore the optimal design of watermark codes, we repeated the deletion experiment
while varying the number of watermark bits and the length of the watermark signal (\figref{WatPlot}).
Empirically, it appears from this preliminary experiment that the single most influential factor on code accuracy is the
correct recovery of LDPC block boundaries, since missed boundaries cause large chunks of the signal
to be dropped, and if not dropped then they can still be impossible to error-correct if the block alignment is wrong.
The single factor that seems most important for correct block boundary identification and alignment
is the ratio of signal bits to watermark bits: the sparser the signal relative to the watermark,
the better chance the decoder has of recovering from large and/or frequent deletions.

\section{Discussion}
\seclabel{Discussion}

We have outlined general design principles for efficiently encoding signals in DNA
by converting a binary sequence to a mixed-radix sequence and then a nonrepeating nucleotide sequence.
A distinct feature of our approach is its modularity,
which allows integrated simultaneous decoding of multi-layer codes that are representable as finite-state transducers.
This allows for straightforward combination and refinement of component automata for
interleaving signals with watermarks, inserting markers or metadata, adding parity bits, converting binary to mixed-radix, generating repeat-free DNA,
modeling errors, or other tasks.

An important aspect of DNA storage which we have not explicitly addressed
is the higher-level organization of the DNA ``filesystem'',
including aspects such as block or packet structure, redundancy, addressing,
versioning, and metadata.
This has been explored in work by others \cite{YazdiEtAl2015,BornholtEtAl2016}.
The code described here operates at a lower level.
However, our allowance for reserved ``control words''
can be used to embed in-band metadata that may be useful in organizing files into manageably-sized but readily-reassembled blocks or packets.
Given that assembly of shorter DNA synthesis into longer contiguous sequences
is currently one of the more expensive and tricky parts of DNA synthesis \cite{GibsonEtAl2009},
and indeed one of the more algorithmically challenging steps of DNA sequencing \cite{PevznerEtAl2001},
and given the difficulties we observe in maintaining code synchronization over longer sequences,
the optimal structure for a DNA filesystem may be one in which a large file is broken into many short
fragments, each individually addressed, which may be reassembled by the decoder, somewhat analogously to IP packets.
In this case nucleic acid control words flanking encoded packet addresses could theoretically be recognized and selected by
molecular hybridization.

Alternatively, since pulling out targeted fragments in this way is difficult and unreliable,
it may be better to encode the fragments in such a way that each fragment combines information
from multiple blocks. This can be done in such a way that a sufficiently large sample of them
is asymptotically likely to guarantee exact recovery of the entire file,
with the asymptotic convergence occurring more rapidly than if each packet simply encodes a single block.
For example, one could construct each fragment by XOR'ing a (quasi)random selection of $k$ blocks from the file,
in the manner of fountain codes \cite{Luby2002,Mackay2005fountain}.
A simpler XOR'ing strategy was partially pursued by Bornholt {\em et al},
who XOR'd pairs of blocks in their DNA storage system \cite{BornholtEtAl2016},
but it can be extended to larger numbers of blocks for more rapid recovery of the entire file
at the expense of a moderate increase in the compute power of the decoder \cite{Mackay2003,Mackay2005fountain}.
The distribution of $k$ can be chosen to optimize decoder convergence.
Results from fountain codes suggest, in practice, the total size of fragments that need to be sequenced exceeds
the file size by only around 5\% \cite{Mackay2003}.

The decoding approach that we have used (Viterbi decoding, followed by decoding of any outer codes such as LDPC) is rather simple.
It does not use the posterior probability information available from the HMM, or allow any other cross-talk between the (cyclic) LDPC factor graph and the (trellis) HMM graph.
It also does not incorporate multiple reads from the same DNA sequence,
and it is subject to computational resource constraints due to the number of states and transitions in the composite transducer,
which grow quickly when multiple coding layers are composed on top of one another.
These are all, in principle, features that could be addressed within the probabilistic framework by approximate or stochastic inference algorithms,
such as loopy belief propagation or Markov Chain Monte Carlo.


\newpage
\section{References}
\bibliography{trans}

\newpage
\section{Tables}

\newpage
\begin{table}[h!t]
\begin{tabular}{rrrrrrr}
& & & & \multicolumn{3}{c}{Mean output symbols/input bit} \\
\cline{5-7}
$N$ & $\nu$ & States & Transitions & Radix 2 & Radix 3 & Radix 4 \\
\hline
\input{mixradtab.tex}
\hline
\multicolumn{4}{r}{Theoretical minima:} & $\log_2(2) = 1$ & $\log_3(2) \simeq 0.631$ & $\log_4(2) = 0.5$
\end{tabular}
\caption{
  \tablabel{MIXRADAR.codes}
  Statistics for several code transducers generated using MIXRADAR (\secref{MIXRADAR}).
$N$ is the block length in bits; $\nu$ is the probability of the end-of-block symbol
(signifying premature termination of the codeword) appearing at any position.
}
\end{table}

\newpage
\begin{table}[h!t]
  \begin{tabular}{rrrrrrrll}
    $\kmerlen$ & $\invreplen$ & $\ncontrols$ & Repeats? & States & Transitions & Bases/bit & Control words & Notes \\
    \hline
\input{dnastoretab.tex}
\end{tabular}
\caption{
  \tablabel{DNASTORE.codes}
  Statistics for several code transducers generated using DNASTORE (\secref{DNASTORE}).
  $\kmerlen$ is the codeword length,
  $\invreplen$ is the minimum length for excluding nonlocal exact inverted repeats
  (which must be separated by at least 2 nucleotides),
  and $\ncontrols$ is the number of control words.
  The first six rows correspond to the transducers shown in \figref{DNAStore}.
  In general the expected bases/bit rise with the codeword length $\kmerlen$,
  since more repeats are excluded from the De Bruijn graph.
}
\end{table}

\newpage
\section{Figures}

\newpage
\begin{figure}[h!t]
\begin{tabular}{ll}
(a) \includedot{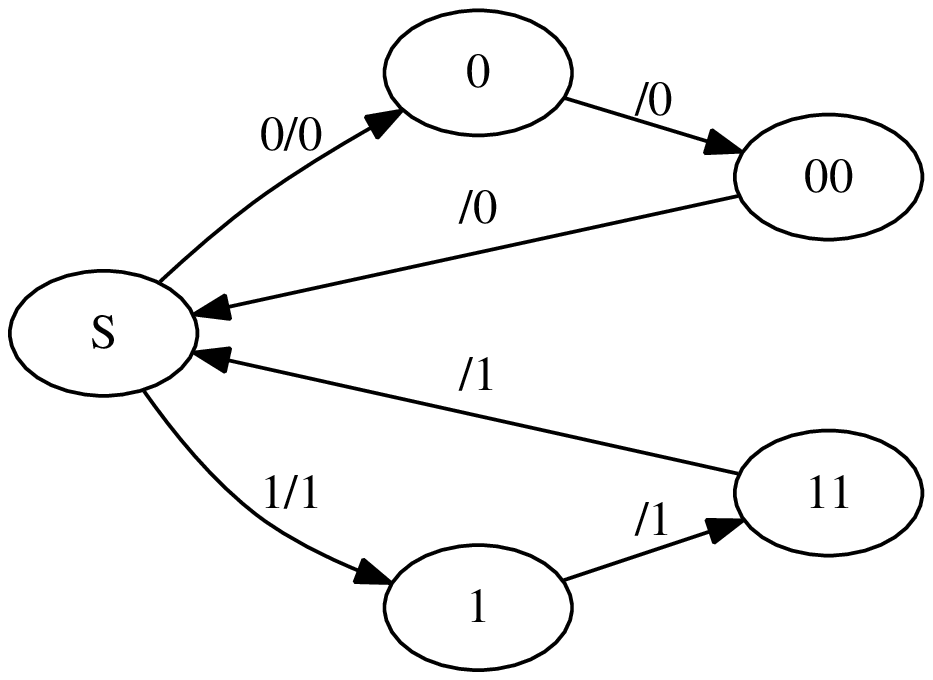}{width=.45\textwidth}
&
(b) \includegraphics[width=.45\textwidth]{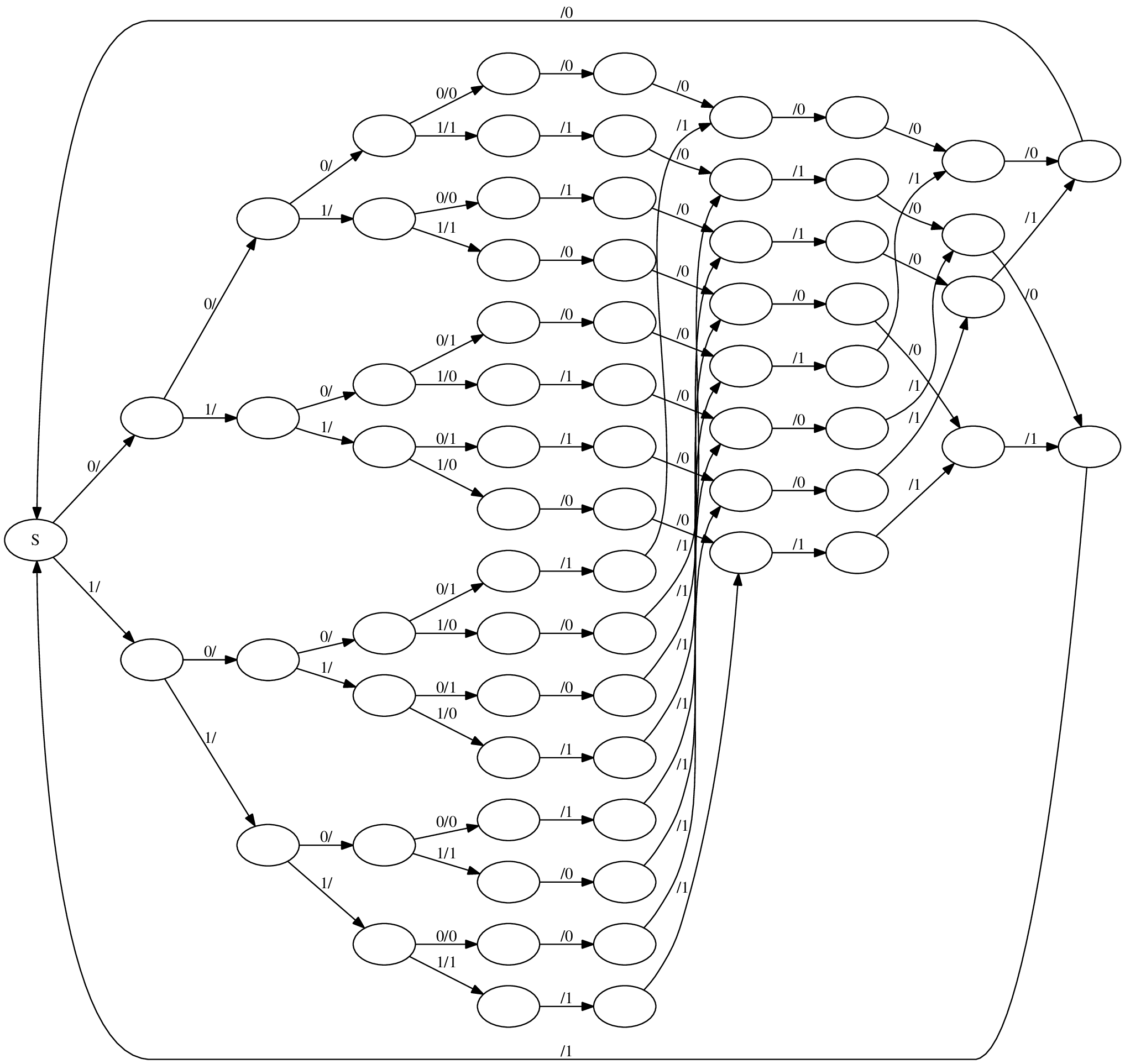}
\end{tabular}
\caption{ \figlabel{HammingTransducer}
State machines implementing Hamming codes for error correction.
Left:
Transducer $\hammingthreeone$ implements the Hamming(3,1) error-correcting code
(which simply repeats every bit three times).
S is both the initial state and the final state.
Right:
Transducer $\hammingsevenfour$ implements the Hamming(7,4) error-correcting code,
with four data bits and three parity bits.
Due to the large number of states, the state names have been omitted from this diagram,
as have the $\epsilon$ labels for empty inputs or outputs.
}
\end{figure}

\newpage
\easyfig{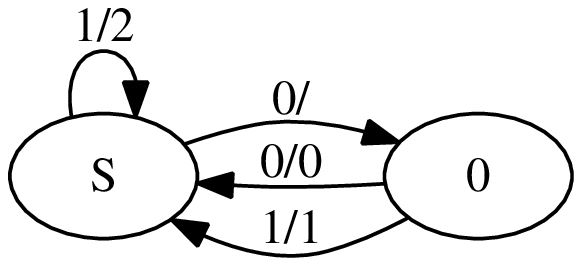}{width=\textwidth}{BinaryToTrinary}{
  A transducer that converts binary to ternary.
  This machine maps $\bit{0}\bit{0} \to \trit{0}$,
  $\bit{0}\bit{1} \to \trit{1}$
  and
  $\bit{1} \to \trit{2}$.
  The problem with this machine is that a dangling zero bit on the input
  can leave it in a state (0)
  that needs to be ``flushed'' with an extra input bit before the machine can finish.
}

\newpage
\easyfig{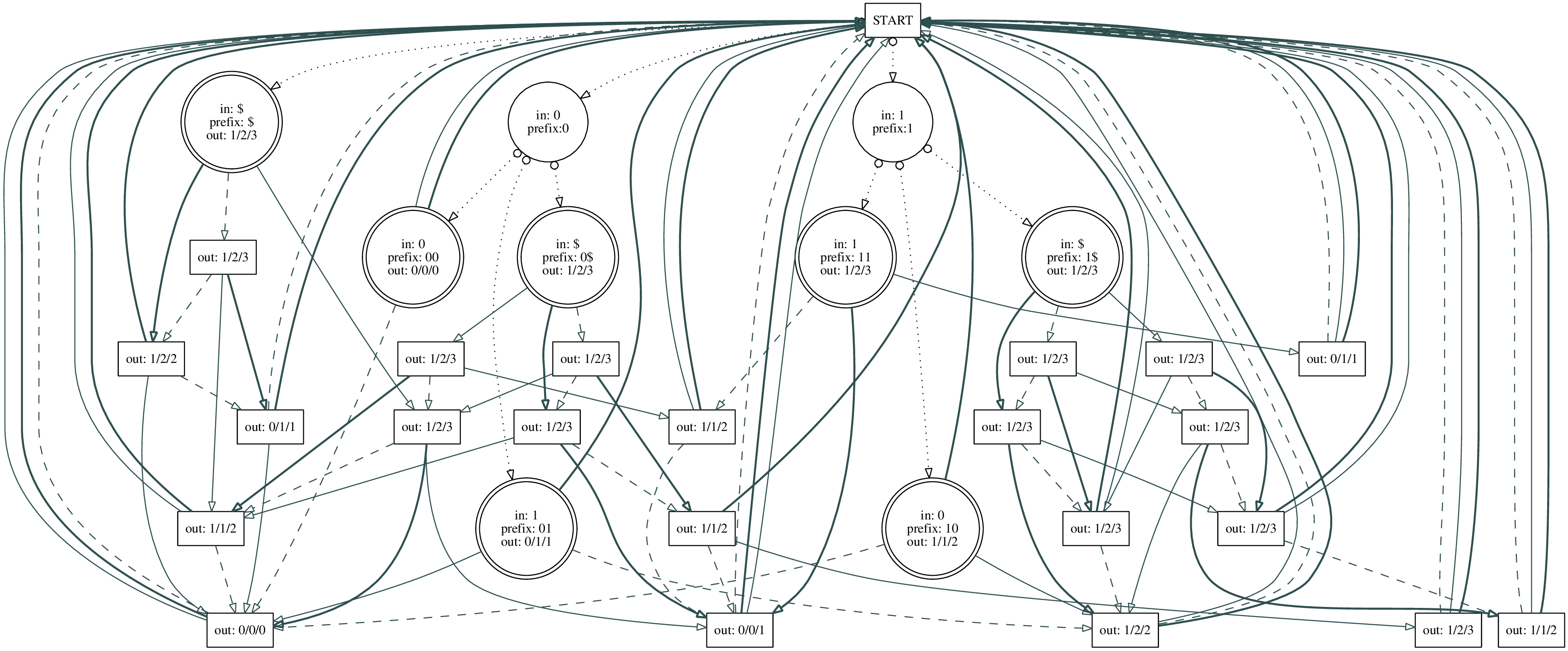}{width=\textwidth}{Arithmetic}{
  A transducer generated by MIXRADAR (\secref{MIXRADAR}) using
  \algref{MixedRadixTransducer} of \secref{Arithmetic}
    with block length 2
    and parameter $\nu$ (the probability of a premature end-of-block) equal to 1/100.
  This transducer converts a binary codeword of length 0, 1 or 2 bits
  into a sequence of mixed-radix selected by the receiver.
  The length of the output sequence in general depends on the radices accepted at each position.
  For example, the input sequence $1_2 1_2$ is converted into the output sequences
  $1_2 1_2 0_2$,
  $1_2 1_2 0_3$,
  $1_2 1_2 1_4$,
  $1_2 1_3$,
  $1_2 1_4$,
  $2_3 0_2$,
  $2_3 1_3$,
  $2_3 1_4$,
  $3_4 0_2$,
  $3_4 0_3$, or
  $3_4 1_4$ (requiring three output symbols if the first two symbols are binary, and two otherwise),
  while the input sequence $0_2 0_2$ is converted into the output sequences
  $0_2 0_2$, $0_2 0_3$, $0_2 0_4$,
  $0_3$, or $0_4$
  (requiring two output symbols if the first symbol is binary, and one otherwise).
  Dotted lines with circles at the source end
  show transitions that input a bit or an end-of-block symbol (`\$').
  Dashed lines show transitions that output a radix-2 digit (i.e. a bit).
  Solid regular-weight lines show transitions that output a radix-3 digit (i.e. a trit).
  Solid bold-weight lines show transitions that output a radix-4 digit (i.e. a quat).
  States are labeled with (as applicable) their input symbol,
  the input prefix so far, and the choice of output symbols
  (depending on whether the radix at the next position is 2, 3, or 4).
  Thus, transitions leaving a state labeled ``out: $i$/$j$/$k$'' output symbols $i_2$, $j_3$ and $k_4$.
}

\newpage
\begin{figure}[h!t]
\begin{tabular}{ccc}
(a) \includedot{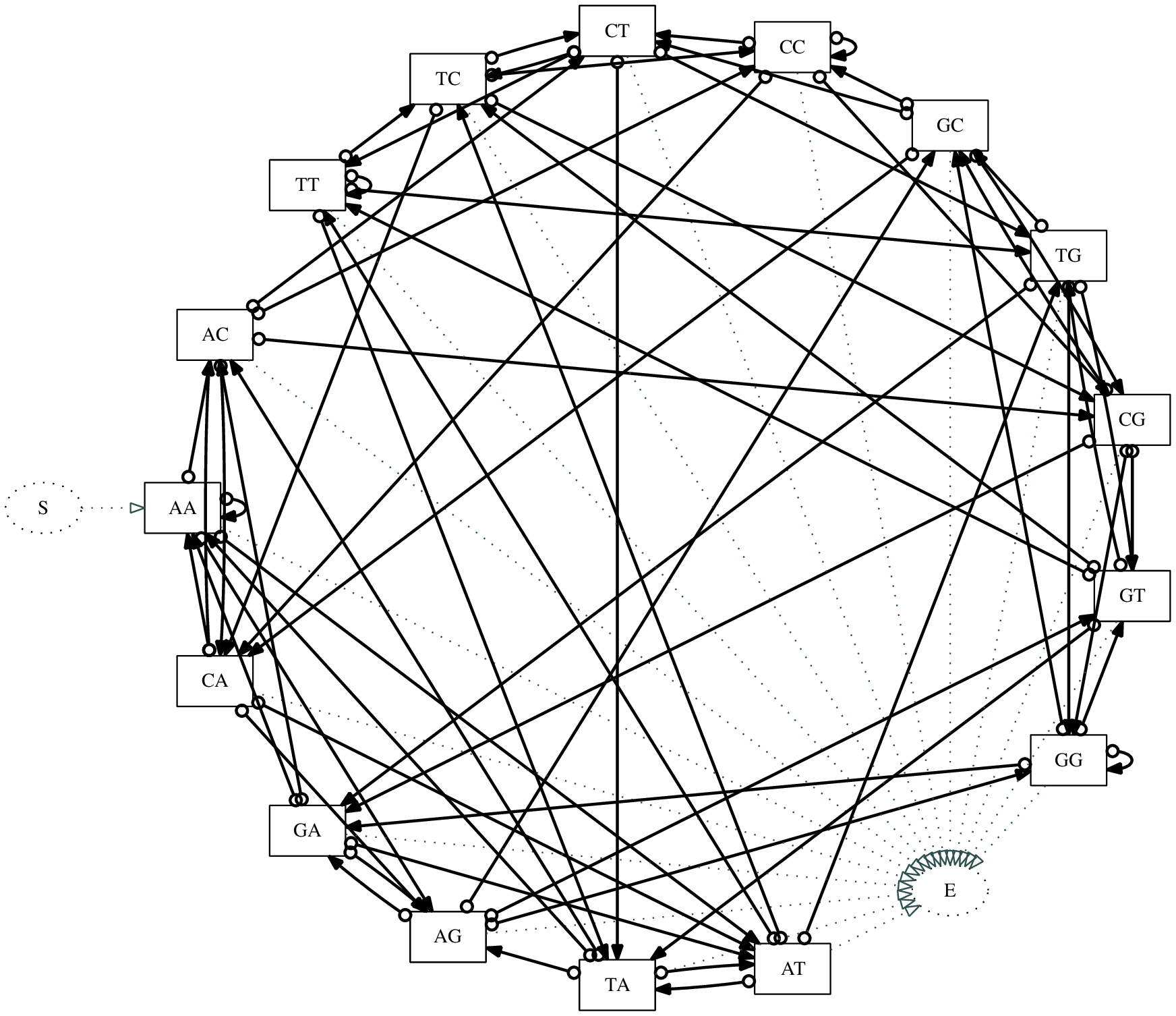}{width=.3\textwidth}
&
(b) \includedot{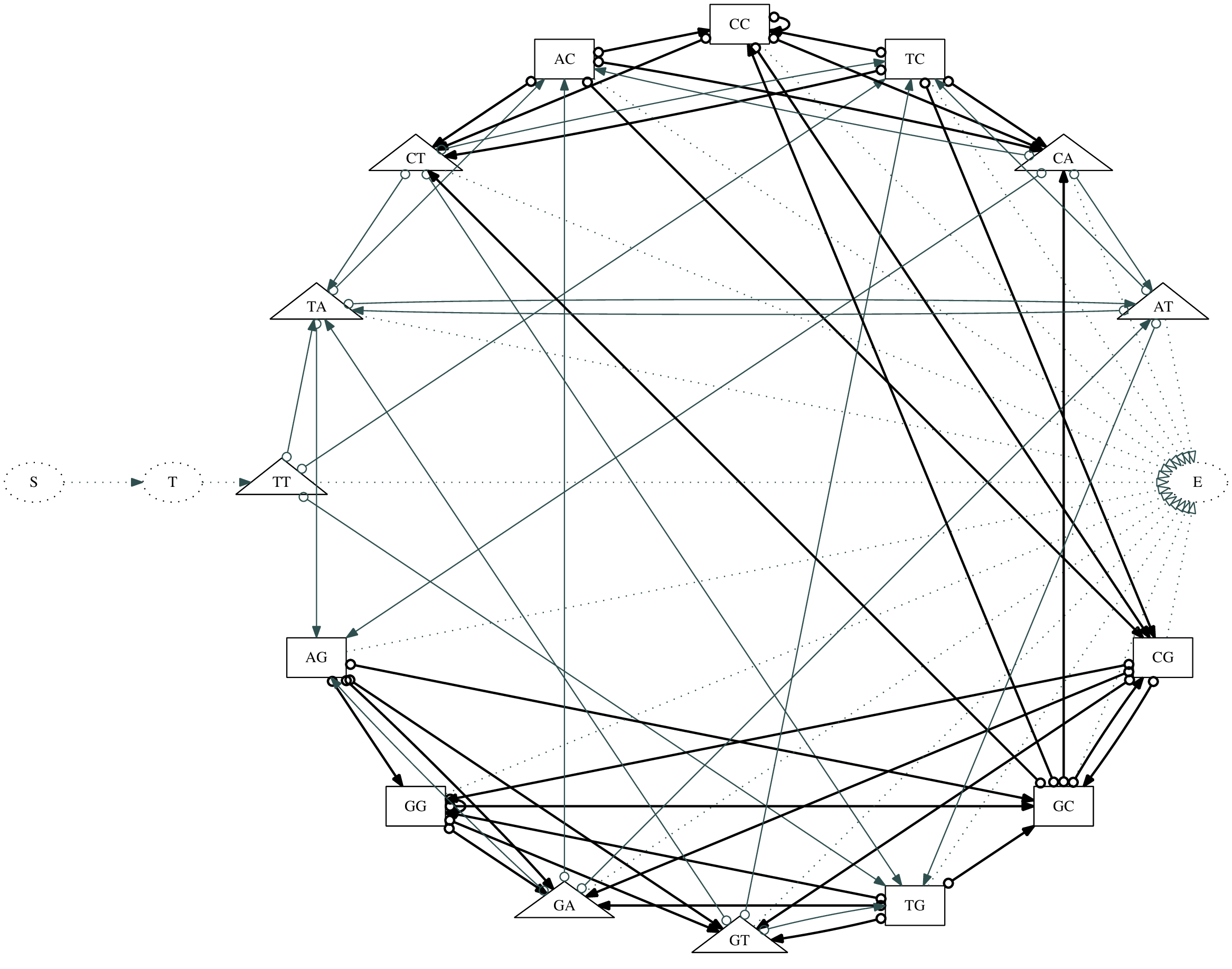}{width=.3\textwidth}
&
(c) \includedot{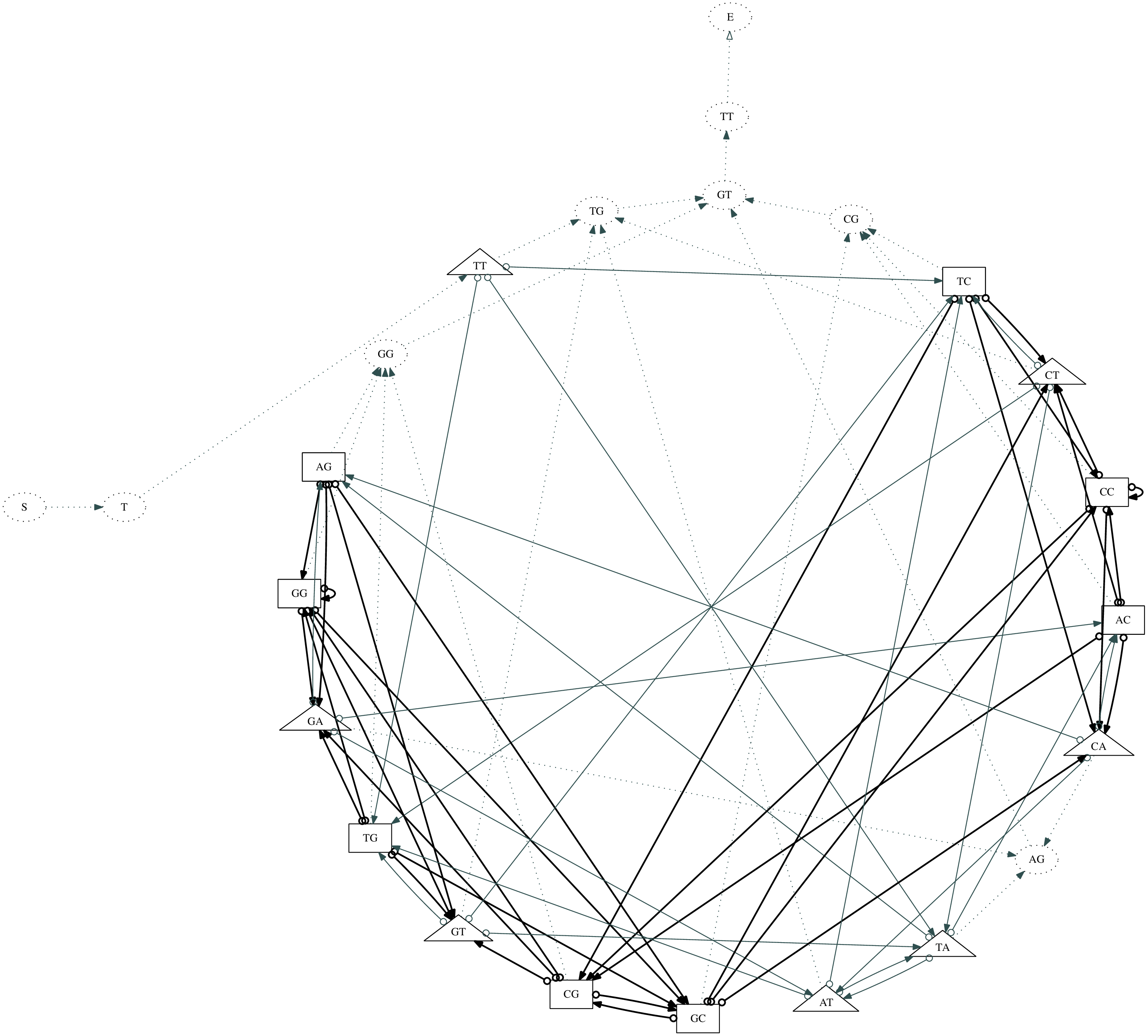}{width=.3\textwidth}
\\
(d) \includedot{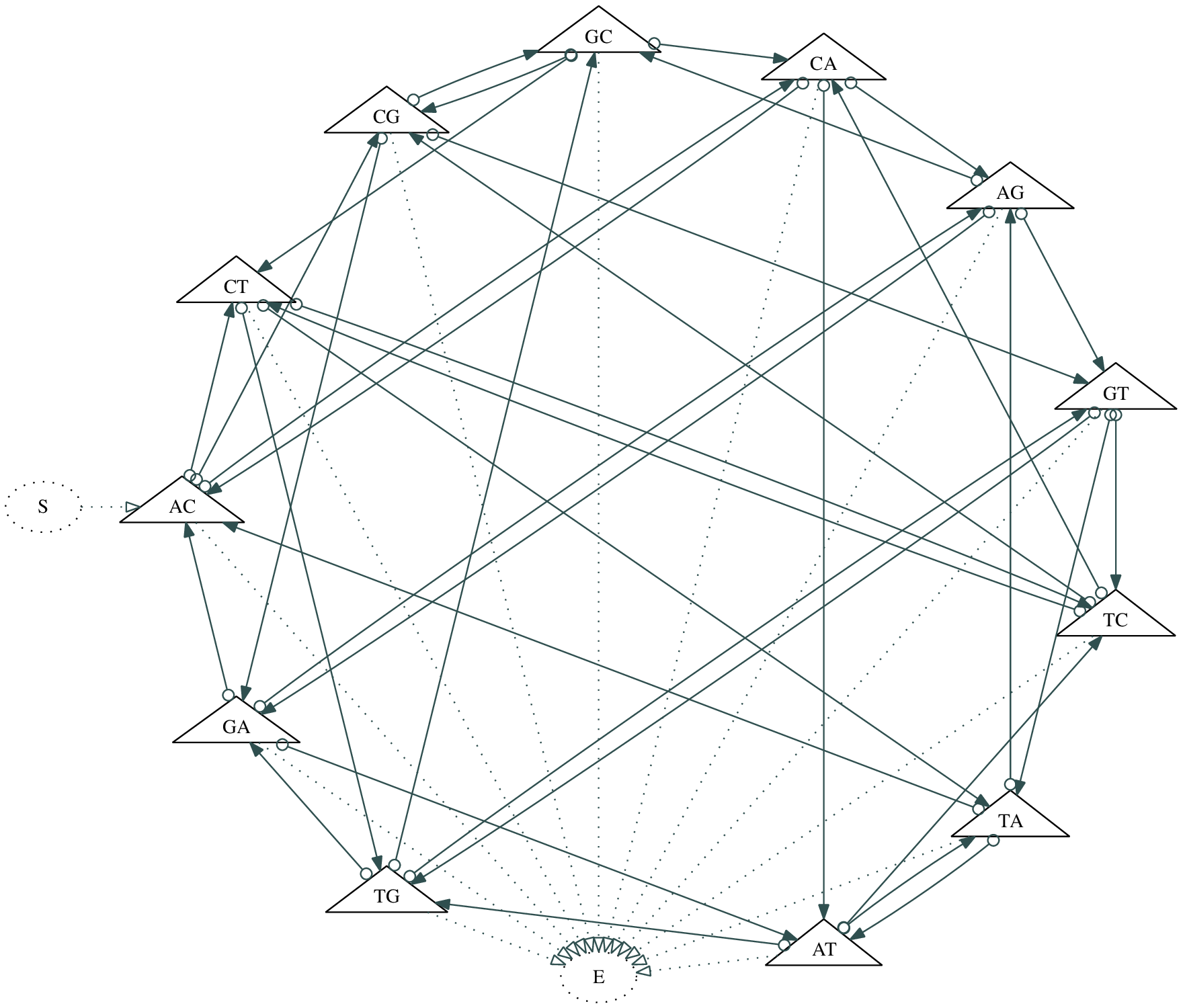}{width=.3\textwidth}
&
(e) \includedot{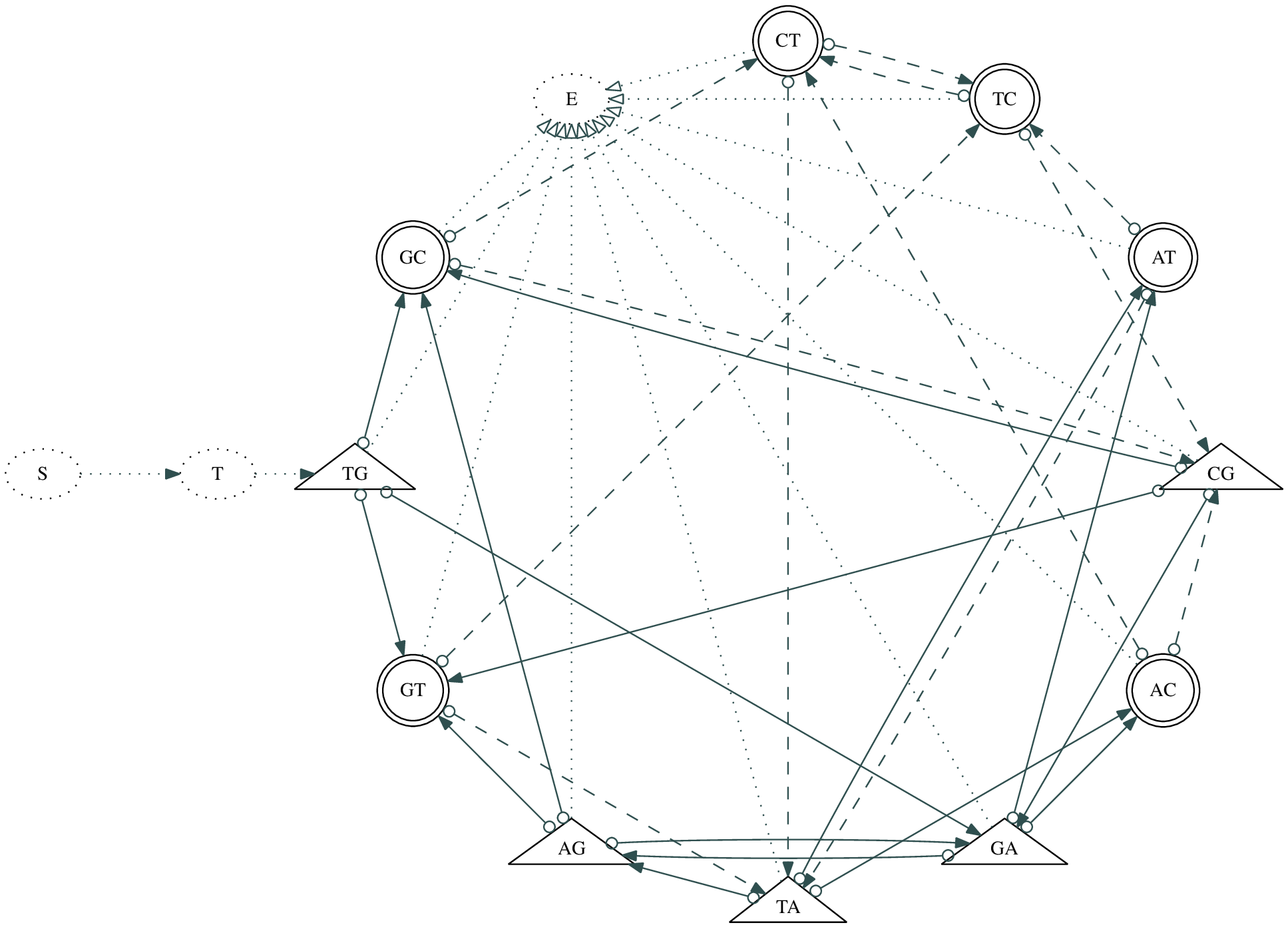}{width=.3\textwidth}
&
(f) \includedot{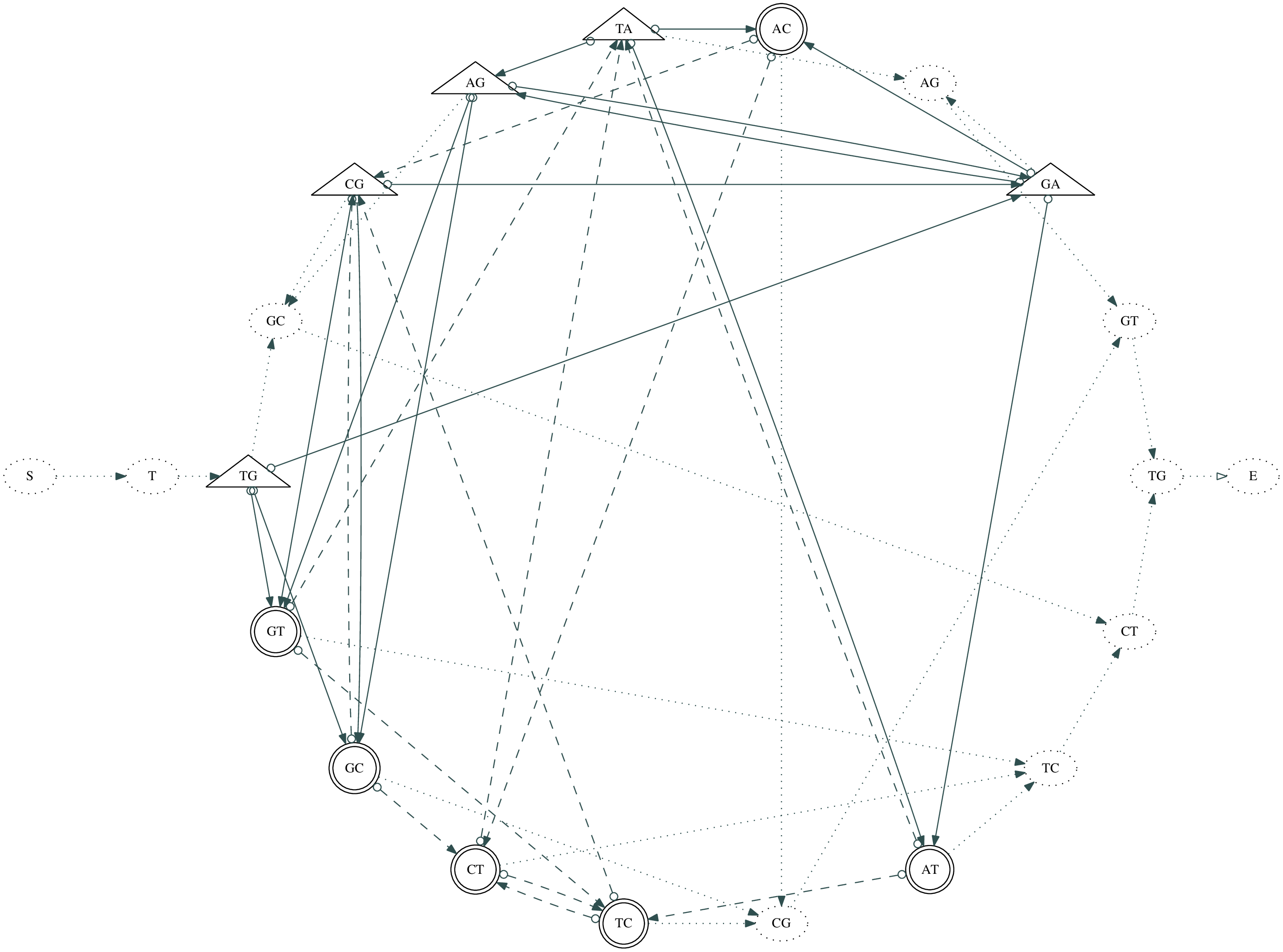}{width=.4\textwidth}
\end{tabular}
\caption{
  \figlabel{DNAStore}
  Transducers generated by DNASTORE (\secref{DNASTORE})
  using the method of \secref{DeBruijnTransducer} with $\kmerlen=2$.
  These codes are all fundamentally based on the 2-dimensional De Bruijn graph
  from which vertices are duplicated, deleted and added to arrive at a transition graph with the required properties.
  Top row (a,b,c): codes in which dinucleotide repeats are allowed.
  Bottom row (d,e,f): codes in which dinucleotide repeats are prohibited.
  Left column (a,d): codes in which there are no reserved control words, so the machine start and end in arbitrary states.
  Central column (b,e): codes in which there is one reserved control word, which only ever appears once, at the start of the encoded DNA sequence.
  Right-hand column (c,f): codes in which there is one reserved control word, which only ever appears twice: once at the start of the encoded DNA sequence and once at the end.
  The leftmost machine in the bottom row (d) is similar to the ternary code of \cite{GoldmanEtAl2013}.
  {\bf Key:}
  Transition label annotations have been omitted from this diagram.
  Instead, the labels may be deduced from the node and edge shapes, as follows:
  Solid bold-weight transitions from rectangular states encode quaternary input digits.
  Solid regular-weight transitions from triangular states encode ternary input digits.
  Dashed-line transitions from double-circle states encode binary input digits.
  Dotted-line transitions do not encode input digits;
  states that can only be exited via these transitions are shown as rectangles.
  Transitions that encode input digits have empty circles at the source end;
  transitions that encode output digits have filled arrowheads at the destination end.
  States are labeled with their past context:
  the output label of a transition into a state $XY$ is always either $Y$ or $\epsilon$.
}
\end{figure}

\newpage
\begin{figure}[h!t]
\begin{tabular}{c}
(a) \includedot{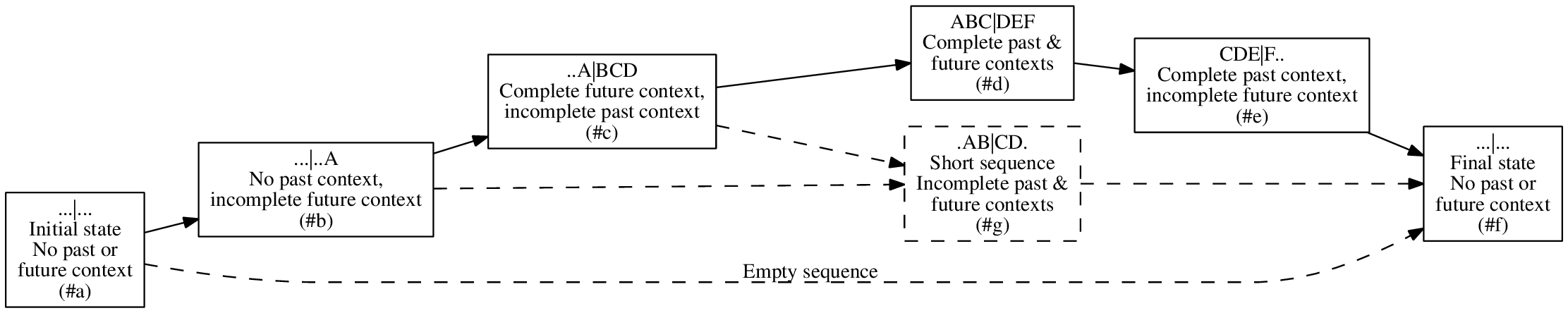}{width=.9\textwidth}
\\
(b) \includedot{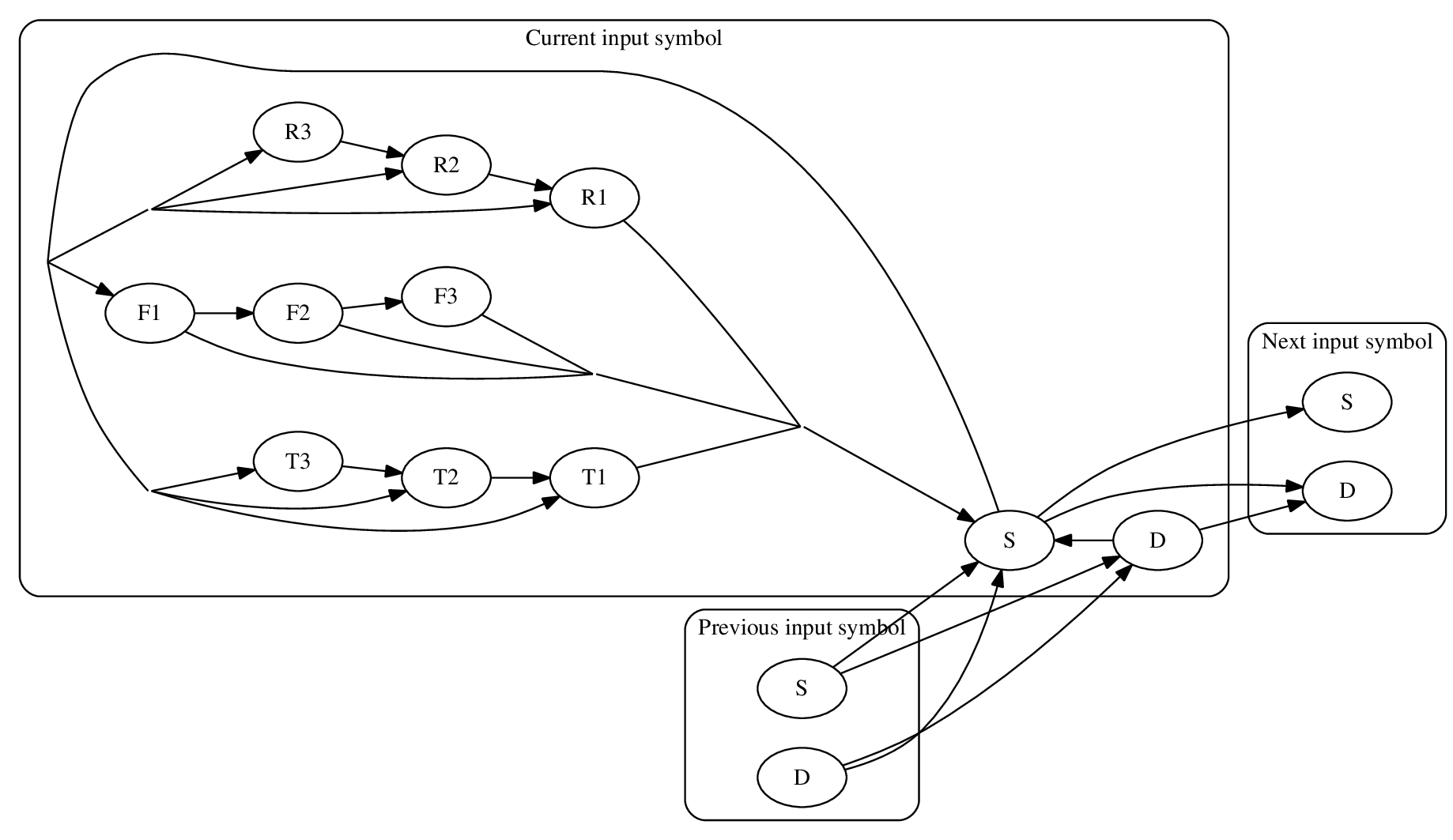}{width=.9\textwidth}
\end{tabular}
\caption{
  \figlabel{PartialObservation}
  Models of error and partial observation (\secref{ErrorModel}).
  Top: Higher-order structure of the error model, showing blocks \#a through \#f.
  The dashed lines show the extra block (\#g) and transitions that would need to be added to handle
  sequences shorter than the total (past+future) context length,
  which require paths that bypass the full loading of both context queues (block \#d).
  Middle: Local neighborhood of a state in the error model for context length $\contextlen=3$.
  For simplicity, the diagram includes only transitions within block \#d, and
  only transitions to or from states with a given context (representing the ``current'' input symbol).
  The $D$ states model deletions;
  the $S$ states model substitutions;
  $T1,T2,T3$ model tandem duplications
  (given past context $ACG$, insert $ACG$, yielding the observed mutation $ACG \to ACGACG$);
  $F1,F2,F3$ model forward inverted duplications
  (given past context $ACG$, insert $CGT$, yielding the observed mutation $ACG \to ACGCGT$);
  and
  $R1,R2,R3$ model reverse inverted duplications
  (given future context $ACG$, insert $CGT$, yielding the observed mutation $ACG \to CGTACG$).
  The probabilities and length distributions of these various events can be modeled.
}
\end{figure}

\newpage
\easyfig{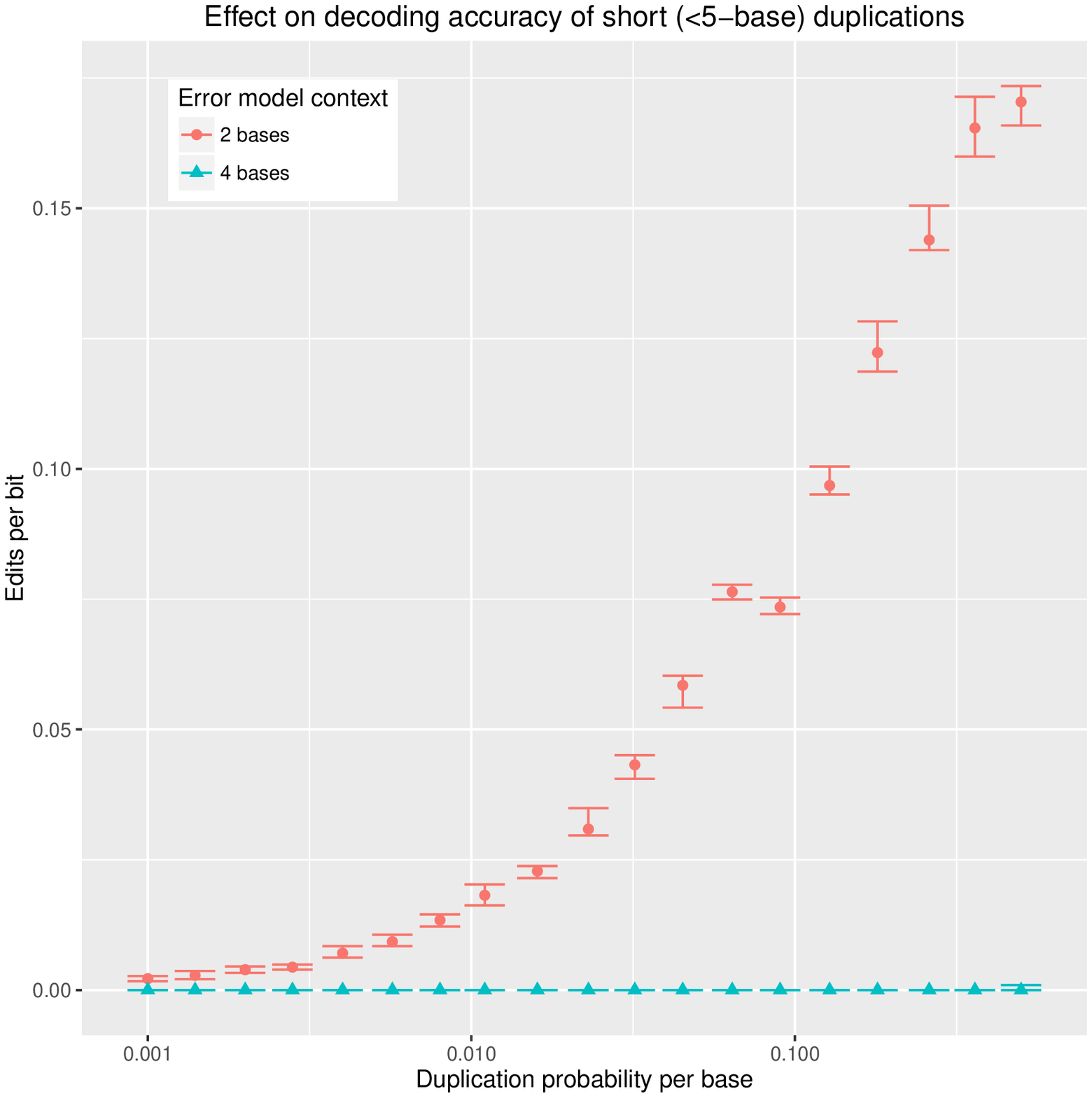}{width=\textwidth}{DupPlot}{
  Effect of simulated duplications on decoder accuracy for two different codes with
  error models of different context length.
  The codes combine the 2-nucleotide mixed-radix arithmetic code of \figref{Arithmetic}
  (MIXRADAR(2))
  with the 4- and 8-nucleotide non-repeating codes of
  \tabref{DNASTORE.codes}
  (DNASTORE(4) and DNASTORE(8)).
  MIXRADAR(2)+DNASTORE(4) has 2 bases of error-model context;
  MIXRADAR(2)+DNASTORE(8) has 4 bases of error-model context.
  The x-axis is the probability of initiating a duplication event at each base;
  from one to four nucleotides were copied in each random duplication event.
  Overlapping (nested) duplications were excluded.
  The y-axis is the Levenshtein edit distance scaled by the length of the input sequence
  (8192 bits). Median and interquartile range are shown.
  The decoder with 2 bases of error context is unable to recognize duplications of 3 or 4
  nucleotides, and so introduces errors during decoding.
}

\newpage
\easyfig{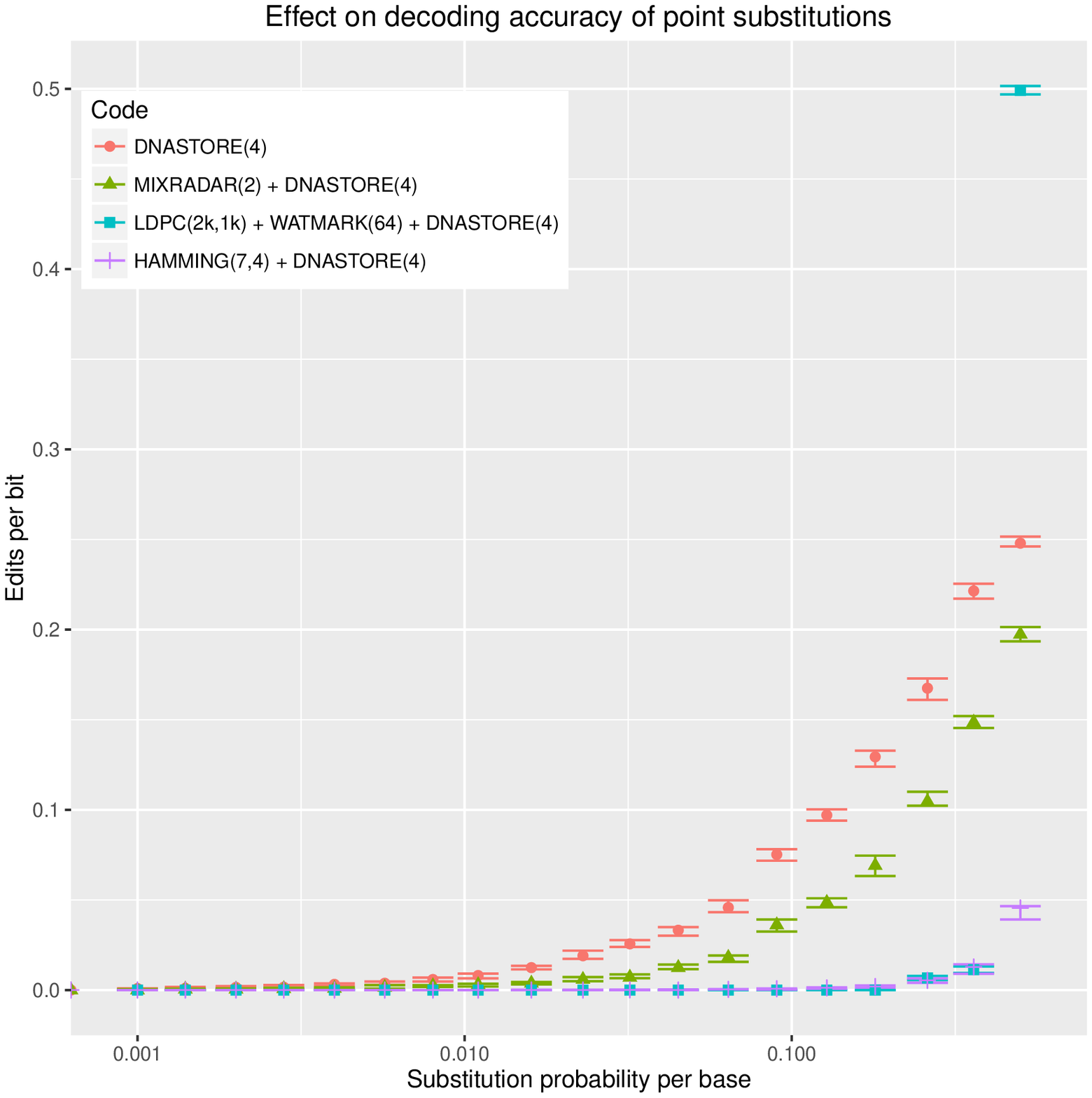}{width=\textwidth}{SubPlot}{
  Effect of simulated substitution errors on decoder accuracy for four different codes.
  The codes all start with the 4-nucleotide non-repeating code of
  \tabref{DNASTORE.codes}
  (DNASTORE(4)).
  On top of this are layered the 2-nucleotide mixed-radix arithmetic code of \figref{Arithmetic}
  (MIXRADAR(2)),
  a 2048-bit, 1024-parity bit low-density parity check code combined with a 64-bit watermark code
  (LDPC(2k,1k)+WATMARK(64)),
  and the Hamming(7,4) code as implemented in \subfigref{HammingTransducer}{b}.
  The x-axis is the probability of substituting each base,
  with a transition/transversion ratio of 10.
  The y-axis is the Levenshtein edit distance scaled by the length of the input sequence
  (8192 bits). Median and interquartile range are shown.
}

\newpage
\easyfig{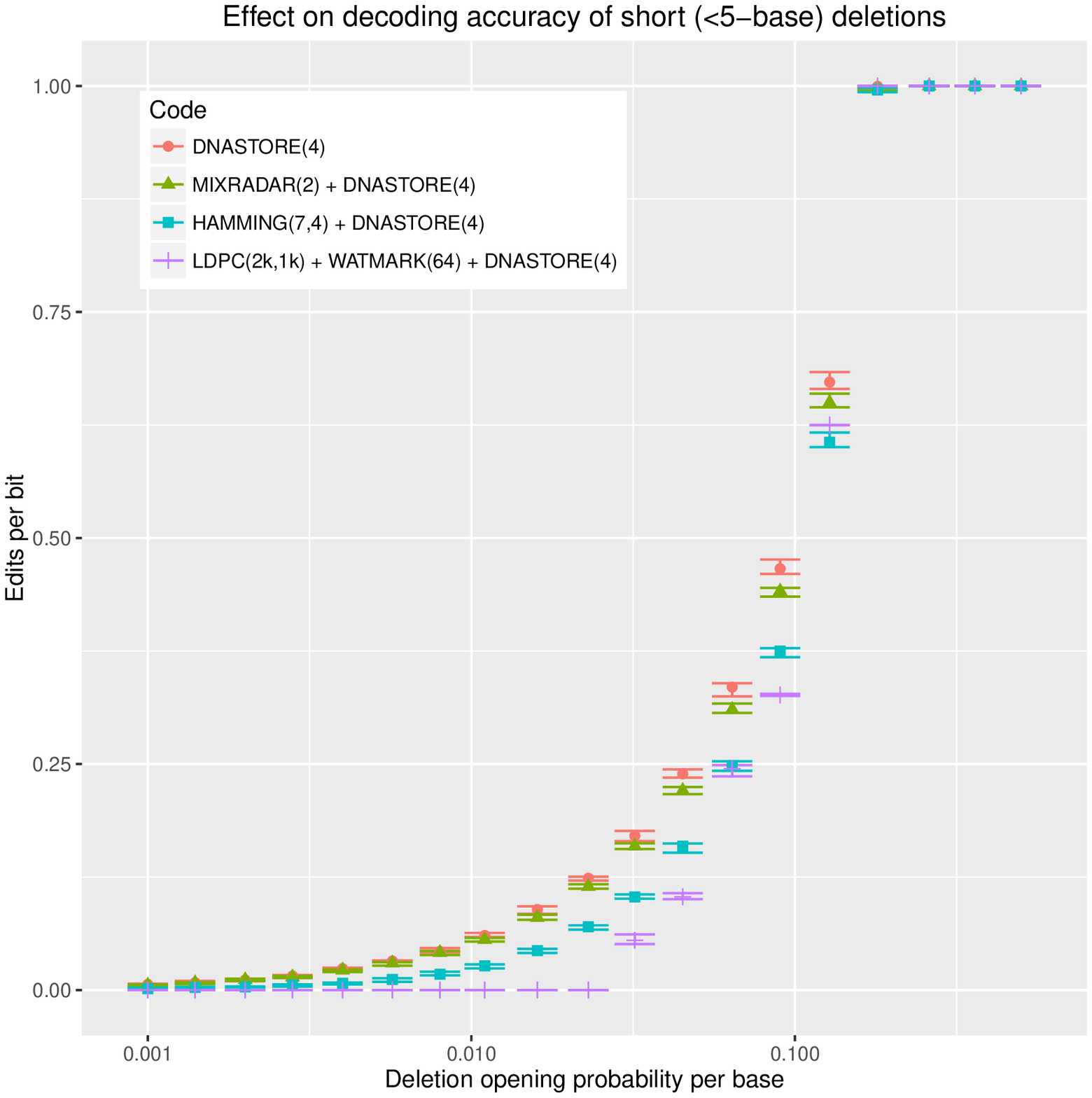}{width=\textwidth}{DelPlot}{
  Effect of simulated deletions on decoder accuracy for four different codes.
  The codes are as described in the caption to \figref{SubPlot}.
  The x-axis is the probability of initiating a deletion event at each base;
  a random number of nucleotides (uniformly sampled from one to four)
  were erased in each deletion event, so that the probability of a nucleotide
  being deleted is $\sim 2.5\times$ the deletion initiation probability.
  The y-axis is the Levenshtein edit distance scaled by the length of the input sequence
  (8192 bits). Median and interquartile range are shown.
}

\newpage
\easyfig{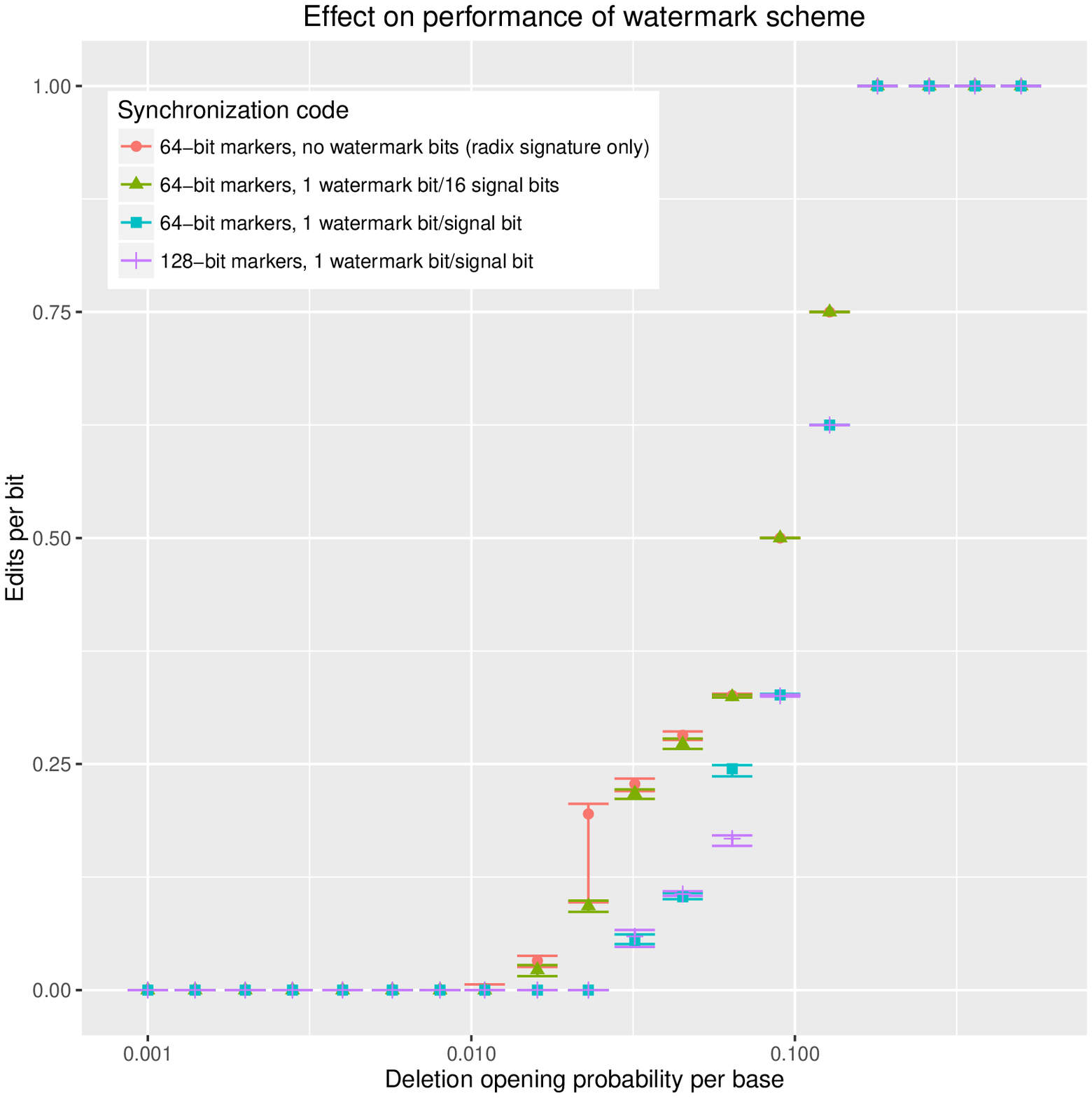}{width=\textwidth}{WatPlot}{
  Effect of watermark scheme on decoder accuracy in presence of simulated deletions.
  The inner code was the 4-nucleotide nonrepeating DNASTORE code
  (\tabref{DNASTORE.codes}).
  Next were various watermarking options:
  (i) 64 bits of signal flanked by a 4-nucleotide control word,
  with a small amount of watermarking information
  encoded in the choice of ternary and quaternary digits used to encode each bit;
  (ii) as (i) but with one pure watermark bit for every 16 signal bits;
  (iii) as (i) but with one pure watermark bit for every signal bit;
  (iv) as (iii) but with a longer period of 128 signal bits
  (and 128 watermark bits) between each control word.
  The outermost code was 2048-bit, 1024-parity bit LDPC.
  The x-axis is the probability of initiating a deletion event at each base;
  the probability of deleting a nucleotide
  is $\sim 2.5\times$ this, as outlined in the caption to \figref{DelPlot}.
  The y-axis is the Levenshtein edit distance scaled by the input sequence length
  (8192 bits). Median and interquartile range are shown.
  The y-axis values are apparently discretized
  because the 8192-bit message is split into eight 1024-bit blocks for LDPC.
  The LDPC decoder discards incomplete blocks,
  so the decoded sequence is close to a multiple of 1024 in length.
  The length difference between the sequences dominates the edit distance:
  once the decoder starts missing whole blocks, the
  per-bit edit distance tends to be rounded up to the next multiple of 1/8.
}

\end{document}

%% file: defs.tex
\newcommand{\secref}[1]{Section~\ref{sec.#1}}
\newcommand{\seclabel}[1]{\label{sec.#1}}

\newcommand{\tabnum}[1]{\ref{tab.#1}}
\newcommand{\tabref}[1]{Table~\tabnum{#1}}
\newcommand{\tablabel}[1]{\label{tab.#1}}

\newcommand{\figlabel}[1]{\label{Figures.#1}}
\newcommand{\figref}[1]{Figure~\ref{Figures.#1}}
\newcommand{\subfigref}[2]{Figure~\ref{Figures.#1}(#2)}
\newcommand{\figrange}[2]{Figures~\ref{Figures.#1} to~\ref{Figures.#2}}

\newcommand{\alglabel}[1]{\label{Algorithms.#1}}
\newcommand{\algref}[1]{Algorithm~\ref{Algorithms.#1}}

\newcommand{\easyfig}[4]{
\begin{figure}[h!t]
\includegraphics[#2]{#1}
\caption{ \figlabel{#3} #4}
\end{figure}}

\newcommand{\includedot}[2]{
\includegraphics[#2]{#1.ps}
}

\usepackage[boxed,noend]{algorithm2e}
\usepackage{graphicx}
\usepackage{psfrag}
\usepackage{url}
\usepackage{amsmath}
\usepackage{amssymb}
\usepackage{color} 
\usepackage{ifthen}
\usepackage{rotating}
\usepackage{multirow}

%% file: notation.tex
\newcommand\semiring{K}
\newcommand\srset{\mathbb{K}}
\newcommand\srplus{\oplus}
\newcommand\srtimes{\otimes}
\newcommand\srplusid{\underline{0}}
\newcommand\srtimesid{\underline{1}}
\newcommand\srsum{\bigoplus}

\newcommand\inalph{\Sigma}
\newcommand\outalph{\Omega}
\newcommand\states{Q}
\newcommand\transitions{E}
\newcommand\initstate{i}
\newcommand\finalstate{f}
\newcommand\emptystring{\epsilon}
\newcommand\maybe[1]{#1 \cup \{ \emptystring \}}

\newcommand\transducer[1]{#1}
\newcommand\trt{\transducer{T}}
\newcommand\trr{\transducer{R}}
\newcommand\trs{\transducer{S}}
\newcommand\tra{\transducer{A}}
\newcommand\trb{\transducer{B}}

\newcommand\state{q}
\newcommand\trans{t}
\newcommand\src{p}
\newcommand\dest{q}
\newcommand\lab{\ell}
\newcommand\inlab{\lab_i}
\newcommand\outlab{\lab_o}
\newcommand\weight{\mathbb{W}}

\newcommand\tprop[2]{#1_{\transducer{#2}}}
\newcommand\tpt[1]{\tprop{#1}{T}}
\newcommand\tpr[1]{\tprop{#1}{R}}
\newcommand\tps[1]{\tprop{#1}{S}}
\newcommand\tpa[1]{\tprop{#1}{A}}
\newcommand\tpb[1]{\tprop{#1}{B}}

\newcommand\somealph{\Gamma}
\newcommand\someseq{\gamma}
\newcommand\midlab{\lab_m}

\newcommand\transcomp{+}
\newcommand\transconcat{\circ}

\newcommand\inseq{\sigma}
\newcommand\outseq{\omega}

\newcommand\kleene[1]{{#1}^{\ast}}
\newcommand\inseqs{\kleene{\inalph}}
\newcommand\outseqs{\kleene{\outalph}}
\newcommand\someseqs{\kleene{\somealph}}

\newcommand\tfunc[1]{\mathbb{#1}}

\newcommand\bigo{{\cal O}}

\newcommand\seqof[1]{\bar{#1}}
\newcommand\sympast{v}
\newcommand\symnext{w}
\newcommand\seqpast{\seqof{\sympast}}
\newcommand\seqnext{\seqof{\symnext}}

\newcommand\nucalph{\somealph_{\mbox{\tiny DNA}}}
\newcommand\sym{x}
\newcommand\comp[1]{\tilde{#1}}
\newcommand\revcomp[1]{\comp{#1}}
\newcommand\ntrans[1]{\hat{#1}}
\newcommand\seqlen[1]{|#1|}
\newcommand\otherseq{\lambda}

\newcommand\kmerlen{{\cal K}}
\newcommand\invreplen{L_{\mbox{\tiny invrep}}}

\newcommand\graph{{\cal G}}
\newcommand\vertices{{\cal V}}
\newcommand\edges{{\cal E}}

\newcommand\debruijngraph{\graph_b}

\newcommand\debruijnedges{\edges_b}

\newcommand\norepgraph{\graph_r}
\newcommand\norepvertices{\vertices_r}
\newcommand\norepedges{\edges_r}

\newcommand\controlgraph{\graph_c}
\newcommand\controlbridges[1]{\vertices_c(#1)}

\newcommand\prequels{{\cal S}}
\newcommand\stepsto{{\cal L}}

\newcommand\ncontrols{N_c}
\newcommand\controlset{{\cal M}}
\newcommand\controlword{M}

\newcommand\initword{\controlword_i}
\newcommand\finalword{\controlword_f}

\newcommand\initvertex[1]{u^{(#1)}_i}

\newcommand\transcoder{\namedtrans{\mbox{\tiny code}}}
\newcommand\transrad{\namedtrans{\mbox{\tiny radix}}}
\newcommand\transdelay{\namedtrans{\mbox{\tiny delay}}}
\newcommand\transerror{\namedtrans{\mbox{\tiny error}}}

\newcommand\bit[1]{#1_2}
\newcommand\trit[1]{#1_3}
\newcommand\quat[1]{#1_4}
\newcommand\controlsym[1]{M[{#1}]}

\newcommand\namedtrans[1]{\trt_{#1}}
\newcommand\figtrans[2]{\namedtrans{\ref{Figures.#1}#2}}

\newcommand\hammingthreeone{\figtrans{HammingTransducer}{a}}
\newcommand\hammingsevenfour{\figtrans{HammingTransducer}{b}}

\newcommand\nuc[1]{\mbox{#1}}
\newcommand\nuca{\nuc{A}}
\newcommand\nucc{\nuc{C}}
\newcommand\nucg{\nuc{G}}
\newcommand\nuct{\nuc{T}}

\newcommand\contextlen{L}

\newcommand\seqpastsyms[2]{\sympast_{#1} \ldots \sympast_{#2}}
\newcommand\seqnextsyms[2]{\symnext_{#1} \ldots \symnext_{#2}}
\newcommand\outsym{y}

\newcommand\param[1]{\mathbb{P}_{#1}}

\newcommand\pnogap{\param{n}}
\newcommand\pdelopen{\param{d}}
\newcommand\ptandup{\param{t}}
\newcommand\pfwddup{\param{f}}
\newcommand\prevdup{\param{r}}

\newcommand\pdelext{\param{x}}
\newcommand\pdelend{\param{e}}

\newcommand\pmatch{\param{m}}
\newcommand\ptransition{\param{i}}
\newcommand\ptransversion{\param{v}}

\newcommand\submat{\param{s}}
\newcommand\lendist{\param{l}}

\newcommand\psub{\submat(\sym,\outsym)}
\newcommand\psubnext[1]{\submat(\symnext_{#1},\outsym)}
\newcommand\psubpast[1]{\submat(\sympast_{#1},\outsym)}
\newcommand\psubcomppast[1]{\submat(\sympast_{#1},\comp{\outsym})}
\newcommand\psubcompnext[1]{\submat(\symnext_{#1},\comp{\outsym})}

\newcommand\plen[1]{\lendist(#1)}

\newcommand\pcont{\pnogap^\ast}

\newcommand\block{\alpha}
\newcommand\srcblock{\block}
\newcommand\destblock{\beta}

\newcommand\blockstate[2]{{#1}_{#2}}
\newcommand\lenstate[3]{\blockstate{#1}{#2}^{(#3)}}

\newcommand\symstate[1]{\blockstate{S}{#1}}
\newcommand\delstate[1]{\blockstate{D}{#1}}
\newcommand\tanstate[2]{\lenstate{T}{#1}{#2}}
\newcommand\fwdstate[2]{\lenstate{F}{#1}{#2}}
\newcommand\revstate[2]{\lenstate{R}{#1}{#2}}

%% file: mixradtab.tex
2 & 0.01 & 32 & 96 & 1.125 & 0.875 & 0.625 \\
3 & 0.01 & 102 & 306 & 1.042 & 0.6667 & 0.6667 \\
4 & 0.001 & 287 & 861 & 1.016 & 0.7344 & 0.5156 \\
5 & 0.001 & 865 & 2595 & 1.006 & 0.6687 & 0.6 \\
6 & 0.001 & 2125 & 6375 & 1.003 & 0.6667 & 0.5026 \\

%% file: dnastoretab.tex
2 &  & 0 & yes & 50 & 96 & 0.5 &  & \subfigref{DNAStore}{a} \\
2 &  & 1 & yes & 41 & 77 & 0.5607 & {\tt TT} (start) & \subfigref{DNAStore}{b} \\
2 &  & 1 & yes & 47 & 83 & 0.5607 & {\tt TT} (start, end) & \subfigref{DNAStore}{c} \\
2 &  & 0 & no & 26 & 48 & 0.6667 &  & \subfigref{DNAStore}{d} \\
2 &  & 1 & no & 19 & 33 & 0.8298 & {\tt TG} (start) & \subfigref{DNAStore}{e} \\
2 &  & 1 & no & 26 & 40 & 0.8298 & {\tt TG} (start, end) & \subfigref{DNAStore}{f} \\
4 &  & 0 & no & 130 & 256 & 0.8055 &  &  \\
4 &  & 2 & no & 208 & 322 & 0.8377 & {\tt TGTC} (start), {\tt CTGT} (end) &  \\
6 &  & 0 & no & 698 & 1384 & 0.8374 &  &  \\
6 &  & 4 & no & 2033 & 2692 & 0.8597 & {\tt TGTCTG} (start), &  \\
  &  &   &    &      &      &        & {\tt ACAGAC} (end), &  \\
  &  &   &    &      &      &        & {\tt GCGTAG}, &  \\
  &  &   &    &      &      &        & {\tt GTAGCA}  &  \\
8 &  & 0 & no & 3802 & 7512 & 0.8537 &  &  \\
8 &  & 4 & no & 10772 & 14456 & 0.8592 & {\tt TGTCTGTA} (start), &  \\
  &  &   &    &       &       &        & {\tt GTATCTGT} (end), &  \\
  &  &   &    &       &       &        & {\tt CGCTACTC}, &  \\
  &  &   &    &       &       &        & {\tt ACGAGCGT}  &  \\
10 & 4 & 0 & no & 20346 & 39976 & 0.8696 &  &  \\
10 & 4 & 4 & no & 56884 & 76494 & 0.8706 & {\tt TGTCTGTATG} (start), &  \\
   &   &   &    &       &       &        & {\tt ACAGACATAC} (end), &  \\
   &   &   &    &       &       &        & {\tt CACGAGTCGT}, &  \\
   &   &   &    &       &       &        & {\tt GTGCTCAGCA}  &  \\
12 & 4 & 0 & no & 101274 & 196472 & 0.9022 &  &  \\
12 & 4 & 4 & no & 298175 & 393352 & 0.9028 & {\tt TGTCTGTATGTC} (start), &  \\
   &   &   &    &        &        &        & {\tt ACAGATGTCTAT} (end), &  \\
   &   &   &    &        &        &        & {\tt CAGTGCTGACAG}, &  \\
   &   &   &    &        &        &        & {\tt TAGCGTGCGAGC}  &  \\